\documentclass[a4paper,11pt]{article}

\usepackage{jcappub} 

\usepackage[T1]{fontenc} 
\usepackage{natbib}
\usepackage{amssymb}
\usepackage{amsmath}
\usepackage{amsfonts}
\usepackage{graphicx}
\usepackage[hang,tight,raggedright]{subfigure}
\usepackage{multirow}
\usepackage{subfigure}
\usepackage{float}
\usepackage{graphics}
\usepackage{slashed}
\usepackage{color}
\usepackage{bm}
\usepackage{braket}
\usepackage{psfrag}
\usepackage{booktabs}
\usepackage{latexsym}
\usepackage{pythonhighlight}

\usepackage{setspace}
\usepackage{amsmath}
\usepackage{amsthm}
\usepackage{mathrsfs}
\usepackage{epsf}
\usepackage{epsfig}
\usepackage{bbm}
\usepackage{bm}
\usepackage{amsfonts}
\usepackage{amssymb}
\usepackage{latexsym}
\usepackage[english]{babel}
\usepackage{multirow}
\usepackage{float}
\usepackage{url}
\usepackage{slashed}
\usepackage{xcolor} 
\usepackage[utf8]{inputenc}
\usepackage{stmaryrd} 
\usepackage{enumitem}
\usepackage{siunitx}
\usepackage{verbatim}
\usepackage{bbm}

\usepackage{caption3}
\usepackage{appendix}

\makeatletter

\newcommand{\Rmnum}[1]{\expandafter\@slowromancap\romannumeral #1@}
\makeatother

\newcommand{\be}{\begin{equation}}
	\newcommand{\ee}{\end{equation}}
\newcommand{\ba}{\begin{array}}
	\newcommand{\ea}{\end{array}}
\newcommand{\bea}{\begin{eqnarray}}
	\newcommand{\eea}{\end{eqnarray}}

\usepackage{theorem}
{
	\theoremheaderfont{\bfseries}
	\theorembodyfont{\normalfont}
}

\title{Detectability of the Phase Transition Gravitational Waves in the DFSZ axion Model}
\author[a]{Aidi Yang,}
\author[a,1]{and Fa Peng Huang\note{Corresponding author.}}
\emailAdd{huangfp8@sysu.edu.cn}
	
\affiliation[a]{MOE Key Laboratory of TianQin Mission, TianQin Research Center for Gravitational Physics \& School of Physics and Astronomy, Frontiers Science Center for TianQin, Gravitational Wave Research Center of CNSA, Sun Yat-sen University (Zhuhai Campus), Zhuhai 519082, China}

\abstract{
In recent years, an increasing number of studies have focused on using gravitational waves to explore axions and the dynamics of Peccei-Quinn symmetry breaking at high energy scales in the early universe.
To accurately quantify the capability of specific gravitational wave experiments to probe the axion properties, it is crucial to perform precise calculations of gravitational wave signals based on given axion models and to conduct detailed detectability analysis tailored to the experimental configurations. Therefore, in this work, we consider the widely-studied DFSZ axion model and, for the first time, perform precise calculations of the phase transition dynamics parameters and associated gravitational wave signals. Our results demonstrate that the DFSZ model allows a strong first-order phase transition for the Peccei-Quinn symmetry-breaking process at high energy scales exceeding $10^{9}~\mathrm{GeV}$. Moreover, by calculating the signal-to-noise ratio of the gravitational waves and comparing it with the thresholds of the Cosmic Explorer detector, we find that these signals are observable by the Cosmic Explorer with the energy scale range from $10^9~\mathrm{GeV}$ to $10^{12}~\mathrm{GeV}$. Notably, through Fisher Matrix analysis, we find that if Cosmic Explorer detectors observe these gravitational waves, the bubble wall velocity will be the first parameter to be determined. This study demonstrates that gravitational wave detection offers a powerful approach to investigating axion dynamics complementary to other experiments.}


\begin{document}
\maketitle
\flushbottom

\section{Introduction}

The strong CP problem in the Standard Model and the microscopic nature of dark matter are long-standing fundamental problems in particle physics and cosmology. The axion particle~\cite{Peccei:1977hh,Peccei:1977ur,Weinberg:1977ma,Wilczek:1977pj,Kim:1979if,Shifman:1979if,Zhitnitsky:1980tq,Dine:1981rt,Sikivie:2020zpn} can naturally solve the strong CP problem and also be a promising dark matter candidate~\cite{Preskill:1982cy,Dine:1982ah, Sikivie:2006ni, Marsh:2015xka}. The energy scale of the $U(1)$ Peccei-Quinn (PQ) symmetry-breaking process determines the axion mass and its couplings with the Standard Model particles. Therefore, it is crucial to investigate the PQ symmetry-breaking scale and its associated dynamics through experimental approaches. However, detecting axion dynamics in high-energy processes poses significant challenges for colliders and other traditional experiments. Motivated by the great achievements of gravitational wave (GW) observations, we consider the possible GW detection of the dynamical process of $U(1)$ PQ symmetry breaking. GW detection offers not only the potential for precise
measurements of invisible axions but also the ability to investigate axion dynamics and estimate their parameters. In the early universe, this process might generate a strong first-order phase transition (SFOPT)~\cite{Croon:2019iuh,Dev:2019njv,DelleRose:2019pgi,Ghoshal:2020vud,VonHarling:2019rgb}. The observable stochastic gravitational wave background (SGWB) induced by a SFOPT has been predicted in previous studies. The study~\cite{Croon:2019iuh} uses the linear sigma model to study the SFOPT, exploring the possibility of observable GWs.  They find that the amplitude of the GW spectrum depends on the mass of the dynamical axion. In certain axion-like particle (ALP) models that can induce a SFOPT, the complementarity between GW observations, particle physics experiments, and astrophysical observations is demonstrated in the study~\cite{Dev:2019njv}. Furthermore, the study~\cite{DelleRose:2019pgi} discusses various PQ symmetry-breaking models and experimental detection prospects for the generated GW signals in detail. The study~\cite{Ghoshal:2020vud} discusses a totally asymptotically free QCD axion model and investigates a region of the parameter space where the PQ symmetry is broken. They find that the PQ phase transition is SFOPT and can produce GW within the reach of future detectors in this region. 

In Ref.~\cite{ VonHarling:2019rgb}, the authors study the interesting PQ phase transition and estimate the GW signals in various well-motivated models, including the Dine-Fischler-Srednicki-Zhitnitsky (DFSZ) \footnote{The DFSZ axion model is one of the representative axion models, and it often serves as a benchmark model in many axion experiments.} axion model~\cite{Dine:1981rt,Zhitnitsky:1980tq}. The simplified bounce action and the corresponding phase transition parameters are used under certain assumptions to estimate the GW signals in Ref.~\cite{ VonHarling:2019rgb}.
To obtain more precise predictions on the parameter space of a SFOPT and the corresponding GW signals, in this work, we precisely calculate the phase transition dynamics and its associated phase transition GW signals from bubble collisions~\cite{Caprini:2015zlo,Huber:2008hg}, sound waves~\cite{Hindmarsh:2016lnk,Konstandin:2017sat,Hindmarsh:2019phv}, and turbulence~\cite{Caprini:2009yp,RoperPol:2019wvy} employing the minimal DFSZ model as a representative example. Moreover, we perform a Fisher matrix (FM) analysis on the GW signals obtained. This model allows for a SFOPT on a broad energy scale ranging from $10^{9}~\mathrm{GeV}$ to $10^{12}~\mathrm{GeV}$. The study~\cite{Preskill:1982cy, Dine:1982ah, Abbott:1982af} shows that when the $U (1)$ PQ symmetry is broken above $10^{12}~\mathrm{GeV}$, it leads to an unacceptably large axion energy density, which is inconsistent with observations. In the absence of these constraints, we find that phase transitions can occur at energy scales ranging from $10^{9}~\mathrm{GeV}$ to $10^{14}~\mathrm{GeV}$. We use a detailed noise model for the Cosmic Explorer(CE)~\cite{LIGOScientific:2016wof,Reitze:2019iox,Hall:2020dps,Hall:2022dik}, enabling a more accurate assessment of its sensitivity to GW signals from the PQ symmetry-breaking process.  By comparing these results with the expected sensitivity curves of the CE and the Einstein Telescope (ET)~\cite{Sathyaprakash:2012jk,Maggiore:2019uih}, we demonstrate that the GW signals from axion models can be detected by CE with a signal-to-noise ratio (SNR) higher than 8. Furthermore, by using FM (FM) analysis~\cite{Fisher:1922saa} to evaluate the precision with which key phase transition parameters can be measured, our study provides deeper insight into the future detection of GW. We find that the bubble wall velocity will be the most strongly constrained parameter to be measured.

For integrity, we briefly review the axion models in Sec.~\ref{sec_axionmd} and the detection principles of SGWB in Sec.~\ref{sec_gwd}. Readers familiar with these two sections can skip them and proceed directly to Sec.~\ref{sec_gwsnr}.
In Sec.~\ref{sec_gwsnr}, we precisely calculate the phase transition GW signals generated by the PQ symmetry-breaking process in the DFSZ model and discuss their detectability at future GW detectors, such as CE and ET. 
Then, we perform a Fisher analysis for the key phase transition parameters at CE in Sec.~\ref{sec_fm}.
A summary is given in Sec.~\ref{sec_sum}. More details on the calculation are shown in the Appendix.

\section{Axion and DFSZ models}\label{sec_axionmd}
\subsection{Strong CP problem and axion solution}
As a $SU(3)_c$ non-Abelian gauge theory in 4-dimensional spacetime, the QCD Lagrangian density contains a CP-violating term originating from the instanton effects
\begin{equation}
	\mathcal{L}_{\mathrm{QCD}} \supset \bar{\theta} \frac{g_s^2}{32 \pi^2} G^{a \mu \nu} \tilde{G}_{a \mu \nu},
\end{equation}
where $ g_s $ is the $SU(3)_c$ gauge coupling constant, $G^{a \mu \nu}$ is the gluon field-strength tensor, and $\tilde{G}_{a \mu \nu} $ is the dual of  $G_{a \mu \nu}$. This CP-violating parameter $\bar{\theta} $ can induce a neutron electric dipole moment of order~\cite{Pospelov:1999ha}
\begin{equation}
	d_n^{(th)} \simeq 2.4(1.0)\times 10^{-16} \bar{\theta} \,\,\text{e~cm}.
\end{equation}
However, current experiments provide an upper bound to 
the electric dipole moment of the neutron as~\cite{Baker:2006ts,Abel:2020pzs} 
\begin{equation}
	|d_n^{(exp)}| < 1.8 \times 10^{-26}\,\text{e~cm}.
\end{equation}
This requires negligible $ \bar{\theta} $, namely,
\begin{equation}
	\bar{\theta} < 10^{-10}.
\end{equation}
The unnaturally small value of $\bar{\theta}$ is known as the strong CP problem in the Standard Model of particle physics.

Axion models provide a natural explanation for the strong CP problem. The strong CP problem can be resolved by introducing a new global $U(1)$ PQ symmetry~\cite{Peccei:1977hh, Peccei:1977ur}. When the $U(1)$ PQ symmetry is spontaneously broken, it gives rise to a new Nambu-Goldstone boson called the axion. The mass and interaction properties of the axion are closely related to the introduced PQ scalar field $\sigma$.  The presence of the axion allows for the replacement of the CP-violating term  $ \bar{\theta} $ with a dynamic field known as the axion field. The interaction between the axion field $a$ and the gluon field takes the form of $a G \tilde{G} / f_a$, where $f_a$ is the decay constant of the axion. The axion field adjusts itself dynamically to a value that effectively cancels out the $ \bar{\theta} $ by settling at the minimum of the axion potential. Through this dynamical mechanism related to the emergence of the axion particle, the strong CP problem is solved.

The original Peccei-Quinn-Weinberg-Wilczek (PQWW) axion model~\cite{Peccei:1977hh, Peccei:1977ur, Weinberg:1977ma, Wilczek:1977pj} assumed that the PQ symmetry breaking happened just at the electroweak energy scale, namely, 
\begin{equation}
	f_a \simeq 246~ \rm GeV.
\end{equation}
It was immediately excluded by experimental data, including beam-dump and LEP experiments results~\cite{Asano:1981nh,Mimasu:2014nea}. Later, two types of invisible axion models were proposed, featuring a significantly higher PQ energy scale, which decreases the axion mass and reduces its couplings to ordinary matter.


\subsection{Invisible axion model}	

One type of invisible axion is the Kim-Shifman-Vainshtein-Zakharov (KSVZ) model, where the exotic quark $Q_E$ and $Q_E^C$ with a new complex gauge-singlet scalar are included~\cite{Kim:1979if,Shifman:1979if,Kim:1986ax}.
In this work, we study the other type of invisible axion model, namely, the well-known DFSZ model~\cite{Dine:1981rt,Zhitnitsky:1980tq}. In this model, all the Standard Model fields are PQ-charged, which is similar to the original PQWW model. Extensions of the DFSZ model with more doublets have been explored in Refs.~\cite{Celis:2014zaa,Diehl:2023uui}. The minimal DFSZ model has two types, both of which consist of two Higgs doublets and a complex scalar, denoted $\sigma$. For the minimal DFSZ-I model~\cite{DiLuzio:2020wdo}, the action is determined by the Yukawa Lagrangian given below
\begin{equation}
	\mathcal{L}_{\mathrm{DFSZ}\text{-} \mathrm{I}}^Y = -Y_U \, \bar{q}_L u_R H_u - Y_D \, \bar{q}_L d_R H_d - Y_E \, \bar{\ell}_L e_R H_d + \text{h.c.},
\end{equation}
where $Y_U$, $Y_D$, and $Y_E$ are Yukawa coupling matrices for the up-type quarks, down-type quarks, and charged leptons, respectively. $q_L$ is the left-handed quark doublet, which includes up and down-quarks.  $u_R$ is up right-handed quarks, $d_R$ is down right-handed quarks, $e_R$ is right-handed charged-lepton singlet, and $\ell_L$ is left-handed lepton doublet. Another option is to introduce the interaction field  \( \tilde{H}_u = i \sigma_2 H_u^* \) in the lepton sector, giving rise to the minimal DFSZ-II variant~\cite{DiLuzio:2020wdo}
\begin{equation}
	\mathcal{L}_{\mathrm{DFSZ}\text{-} \mathrm{II}}^Y = -Y_U \, \bar{q}_L u_R H_u - Y_D \, \bar{q}_L d_R H_d - Y_E \, \bar{\ell}_L e_R \tilde{H}_u + \text{h.c.}.
\end{equation}
Our research focuses on the minimal DFSZ-I model. 
The tree-level potential is~\cite{Dine:1981rt, Ghigna:1992iv,Ahmadvand:2021vxs,DelleRose:2019pgi,VonHarling:2019rgb}
\begin{align}\label{Vtree1}
	\begin{split}
		V_{\text{tree}} &= -\mu_1^2|H_u|^2 - \mu_2^2 |H_d|^2 + \lambda_1|H_u|^4 + \lambda_2 |H_d|^4  +  \lambda_4|H_u^\dagger H_d|^2\\
		&\quad - \mu_3^2|\sigma|^2  + \lambda_3|\sigma|^4 + \lambda_{12}|H_u|^2 |H_d|^2 + \lambda_{13}|\sigma|^2|H_u|^2  \\
		&\quad+ \lambda_{23}|\sigma|^2 |H_d|^2 + \left(\lambda_5\sigma^2\tilde{H}_u^\dagger H_d + \mathrm{h.c.}\right),\;
	\end{split}
\end{align}
where all coefficients in this model are assumed to be real. The transformation of the scalar fields of this model follows the expression below
\begin{equation}
	\sigma \rightarrow e^{i X_\sigma \gamma} \sigma, \quad H_{u} \rightarrow e^{i X_u \gamma} H_{u}, \quad H_{d} \rightarrow e^{i X_d \gamma} H_{d},
\end{equation}
where $X_\sigma, X_u, X_d$ are the PQ charges corresponding to the PQ symmetry transformations of the fields $\sigma, H_{u}$, and $H_{d}$, respectively. $\gamma$ is the parameter of the PQ symmetry transformation, which is a real number.
\begin{equation}
	X_{u}+X_{d}=2 X_\sigma=1.
\end{equation}
This relation ensures that the Lagrangian is invariant under the PQ symmetry transformation.
In the minimal DFSZ-I model, the domain wall number $N_{DW}=6$ leads to the disastrous domain wall problem. However, introducing explicit PQ symmetry-breaking terms can solve this problem~\cite{Zhang:2023gfu}. At high temperatures, this term generates a thermal effective potential for the axion, which unifies the initial value of the axion field in the early universe, thereby avoiding the formation of domain walls. Since the minimal DFSZ-I model itself contains two Higgs doublets, the following explicit breaking term can be considered as
\begin{equation}
	\mathcal{L}_{\text{break}} = \mu_{4} e^{i\delta} \sigma H_u^\dagger H_d + \text{h.c.,}
\end{equation}
where \( \mu_{4} \) is a dimensional parameter and $\delta$ is the phase. After introducing the explicit breaking term, interactions with standard model particles  generate a thermodynamic potential for the axion at high temperatures
\begin{equation}
	V(a, T) \propto T^{n_{a}} \cos\left( \frac{a}{f_a} + \delta \right),
\end{equation}
where \( n_{a} \) is a positive integer. In the selection of explicit breaking terms, the following factors must be taken into account:
1. The explicit breaking term's energy scale should not significantly affect PQ symmetry-breaking dynamics.
2. The explicit breaking term should generate a thermal potential at high temperatures to prevent the formation of domain walls.  This thermal potential ensures that the axion field quickly rolls to a unique field value throughout space, thereby avoiding the issue of the field acquiring different values in different regions, which would lead to the formation of domain walls.
3. An excessively large potential energy for the axion field from the explicit breaking term as the temperature decreases to the QCD scale should be avoided, as it would cause the vacuum expectation value (VEV) of the axion field to deviate from the value required to solve the strong CP problem, consequently introducing an unacceptable level of CP violation. To satisfy the above conditions, the parameters must be appropriately constrained. To provide an adequate thermal potential at high temperatures, $\mu_4$ must be sufficiently large. However, $\mu_4$ cannot be too large since this term will induce a deviation of the axion's VEV from the value necessary to cancel the CP violation parameter, consequently reintroducing CP violation at low temperatures.

After satisfying these constraints, GWs from domain wall decay can be neglected. Therefore, the explicit breaking term only affects GWs from the PQ phase transition through the following factors:
1. Mass Matrices: This term modifies the mass matrices for the scalar fields.
2. Thermal Corrections: This term can affect thermal masses. However, the explicit breaking term only introduces an additional contribution, similar to the $\lambda_5$ or $\mu_1$ parameters, in the mass matrices or thermal corrections. In our parameter space exploration, we have thoroughly explored a wide range of values for these parameters. The parameter space is not expected to be substantially altered. The specific influence of each parameter on the phase transition is elucidated in the following sections.

In the following sections, for brevity, we refer to the minimal DFSZ-I model as the DFSZ model. Both the KSVZ and DFSZ models can solve the strong CP problem by introducing a PQ symmetry that is spontaneously broken at a high scale, leading to the generation of the axion.  The specific coupling strengths and the resulting axion phenomenology differ between these two models. In the minimal KSVZ model, the phase transition is always second order~\cite{DelleRose:2019pgi,VonHarling:2019rgb}. The study~\cite{DelleRose:2019pgi} discusses two possible modifications of the KSVZ model that can lead to SFOPT and observable GW signals.  However, both modifications compromise the simplicity of the minimal KSVZ model. In contrast, in the minimal DFSZ model, the PQ scalar field directly couples to the Standard Model Higgs field. These particles are abundant at the phase transition temperature, and their thermal corrections significantly contribute to the effective potential. Therefore, the DFSZ model is more favorable for generating SFOPT GW signals compared to the KSVZ model.

\subsection{Axion dark matter and axion detection}

Through the misalignment mechanism or other mechanisms, the axion particle can be a natural dark matter candidate~\cite{Preskill:1982cy,Dine:1982ah, Sikivie:2006ni}. There are various experiments to detect the axion particle or axion dark matter, such as helioscope searches~\cite{Sikivie:1983ip,Lazarus:1992ry,Moriyama:1998kd}, haloscopes~\cite{ADMX:2009iij}, and the radio telescope searches~\cite{Pshirkov:2007st,Huang:2018lxq,Hook:2018iia}, see the recent review~\cite{Sikivie:2020zpn} for details.

GW detection might provide new approaches to explore the axion~\cite{VonHarling:2019rgb} and ALP~\cite{Croon:2019iuh,Chiang:2020aui,Dev:2019njv,DelleRose:2019pgi,Ahmadvand:2021vxs,Ghoshal:2020vud,Xie:2022uvp,Yang:2023aak,Yang:2023vwm}. 
To explore the possibility that the DFSZ axion model can generate an observable SGWB, we review the detection principles of the SGWB~\cite{michelson1987,christensen1992,Allen:1997ad} and the sensitivity curves of GW detectors in the next section. This will lay the foundation for calculating the strength of the phase transition GW signals predicted by the DFSZ model and evaluating their detectability in Sec.\ref{sec_gwsnr}.

\section{ Stochastic gravitational wave background detection }\label{sec_gwd}
The PQ symmetry breaking in the DFSZ axion model discussed in the last section might generate a SFOPT, which can produce an observable SGWB. Unlike GW signals from individual astrophysical events, SGWB~\cite{michelson1987,christensen1992,Allen:1997ad} arises from the superposition of GWs emitted by countless sources across the universe, such as phase transition GW. We cannot directly measure the properties of individual SGWB sources.

To detect the SGWB, multiple GW detectors are typically employed to observe the SGWB simultaneously. Extracting the SGWB signal from the data collected by these GW detectors requires long-term integration and cross-correlation analysis. This section briefly summarises the key points for SGWB detection~\cite{Maggiore:2007ulw}.	

\subsection{Signal analysis}
In the actual detection process, the output signal of the detector, $s_{gw}(t)$, consists of the GW signal, $h_{gw}(t)$, and noise, $n_{gw}(t)$
\begin{equation}\label{shn}
	s_{gw}(t) = h_{gw}(t) + n_{gw}(t).
\end{equation}
Here, it is assumed that the noise, $n_{gw}(t)$, is a stationary random process, and its autocorrelation function depends only on the time difference, not the specific time point. The autocorrelation function of the noise, $R(\tau)$, is defined as 
\begin{equation}\label{Rtau}
	R(\tau) = \langle n_{gw}(t) n_{gw}(t + \tau) \rangle = \int_{-\infty}^{\infty} n_{gw}(t) n_{gw}(t + \tau) dt.
\end{equation}
According to the Wiener-Khinchin theorem, the noise power spectral density (PSD) $S_n(f)$ of a stationary random process is defined as the Fourier transform of its $R(\tau)$
\begin{equation}\label{Sn}
	S_n(f) = 2 \int_{-\infty}^{\infty} d\tau\, R(\tau)\, e^{-2 \pi i f \tau}.
\end{equation}
The mean square value of the noise in the time domain is related to the PSD
\begin{equation}\label{mean_square_noise}
	\langle n_{gw}^2(t) \rangle = R(0) = \int_{0}^{\infty} df S_n(f).
\end{equation}
This equation shows that the total power of the noise is equal to the integral of the PSD over all frequencies. The SNR plays a crucial role in extracting the GW signal from the noisy data. We employ the technique of matched filtering to maximize the SNR.   The matched filter optimally enhances the SNR by correlating the detector output with a filter function $K(t)$~\cite{Maggiore:2007ulw}
\begin{equation}\label{eq_sdef}
	\hat{s} = \int_{-\infty}^{\infty} dt s_{gw}(t) K(t).
\end{equation}
The SNR is defined as $S/N$, where $S$ is the expected value of the filtered output $\hat{s}$ when the signal is present, and $N$ is the root mean square value of $\hat{s}$ in the absence of the signal.  Since $\langle n_{gw}(t) \rangle = 0$, $S$ can be calculated as follows
\begin{equation}
	S  =\int_{-\infty}^{\infty} d t\langle s_{gw} (t)\rangle K(t)  =\int_{-\infty}^{\infty} d t h_{gw} (t) K(t) =\int_{-\infty}^{\infty} d f \tilde{h}_{gw}(f) \tilde{K}^*(f).
\end{equation}
The variance of $N^2$ can be obtained by calculating the convolution of $K(t)$ with $R(\tau)$ which is given by Eq.~\eqref{Rtau}
\begin{equation}
	N^2 =\int_{-\infty}^{\infty} d t d t^{\prime} K(t) K\left(t^{\prime}\right) \int_{-\infty}^{\infty} d f \frac{1}{2} S_n(f) e^{2 \pi i f\left(t-t^{\prime}\right)} =\int_{-\infty}^{\infty} d f \frac{1}{2} S_n(f)|\tilde{K}(f)|^2. 
\end{equation}
Therefore, the SNR for a single detector is
\begin{equation}\label{singlesnr}
	\mathrm{SNR}_{\mathrm{single}}=\frac{\int_{-\infty}^{\infty} d f \tilde{h}_{gw}(f) \tilde{K}^*(f)}{\left[\int_{-\infty}^{\infty} d f \frac{1}{2} S_n(f)|\tilde{K}(f)|^2\right]^{1 / 2}},
\end{equation}
where $\tilde{h}_{gw}(f)$ is Fourier transform of $h_{gw}(t)$ and $\tilde{K}(f)$ is Fourier transform of $K(t)$. 

A single detector cannot effectively distinguish between SGWB signals and detector noise because of their random properties. However, by using two detectors and comparing their outputs, the SGWB signal can be extracted~\cite{Allen:1997ad}. This is achieved by calculating the cross-correlation function between the outputs of the two detectors. The differences in the locations and orientations of the two detectors affect the correlation of the GW signals they detect. This correlation can be described by the overlap reduction function $\Gamma(f)$~\cite{Allen:1997ad}, which accounts for the reduction in sensitivity to the SGWB due to the separation and relative alignment of the detectors. The overlap reduction function is defined as
\begin{equation}\label{overlap_reduction}
	\Gamma(f) = \frac{1}{8\pi} \sum_{A=+,\times} \int_{S^2} d\hat{\Omega} e^{i 2\pi f \hat{\Omega}\cdot\Delta\vec{x}} F_1^A(\hat{\Omega}) F_2^A(\hat{\Omega}),
\end{equation}
where $F_i^A(\hat{\Omega})$ is the response function of detector $i$ for GW propagation direction $\hat{\Omega}$ and polarization $A,~\Delta \vec{x}$ is the separation vector between the detectors.

Similar to the case of a single detector, we also need to maximize the SNR for two detectors. To achieve this, a modified form of the matched filtering function $K(t)$, denoted as $Q(t)$, is introduced for the two-detector scenario~\cite{Maggiore:2007ulw}. And $Q(t)$ plays the same role as the $K(t)$ in Eq.~\eqref{singlesnr}
\begin{equation}\label{eq_snropt}
	\mathrm{SNR}_{\mathrm{optimized}}=T_t^{1 / 2} \frac{\int_{-\infty}^{\infty} d f S_h(f) \Gamma(f) \tilde{Q}(f)}{\left[\int_{-\infty}^{\infty} d f|\tilde{Q}(f)|^2 S_{2n}^2(f)\right]^{1 / 2}},
\end{equation}
where $T_t$ is the observation time, $\tilde{Q}(f)$ is the Fourier transform of ${Q}(t)$ and $S_h(f)$ is the PSD of the SGWB. $S_
{2n}(f)=\left[S_{n, 1}(f) S_{n, 2}(f)\right]^{1 / 2}$ is the geometric mean of the noise PSD of the two detectors. The optimization problem that finds the optimal form of $\tilde{Q}(f)$ to maximize the SNR can be solved using the Schwarz inequality, which states that for any two functions $u(f)$ and $v(f)$
\begin{equation}
	\left|\int_{-\infty}^{\infty} d f u(f) v^*(f)\right|^2 \leq \int_{-\infty}^{\infty} d f|u(f)|^2 \int_{-\infty}^{\infty} d f|v(f)|^2.
\end{equation}
To apply the inequality, we can set
\begin{equation}
	u(f) = \frac{S_h(f) \Gamma(f)}{S_{2n}(f)}, \quad v(f) = \tilde{Q}(f) S_{2n}(f).
\end{equation}
Then, we obtain
\begin{equation}\label{opsnr}
	\mathrm{SNR}_{\mathrm{optimized}} = T_t^{1 / 2}\left[\int_{-\infty}^{\infty} d f \frac{S_h^2(f) \Gamma^2(f)}{S_{2n}^2(f)}\right]^{1 / 2}.
\end{equation}

\subsection{Sensitivity curves }
The sensitivity curve of a detector characterizes its ability to detect signals. Three commonly used definitions for sensitivity curves are the characteristic strain, the amplitude spectral density, and the energy density~\cite{Thrane:2013oya}. 

\subsubsection{	The characteristic strain}

The characteristic strain denoted as $ h_c(f) $ can be derived from the metric perturbations $ h_{a b}(t, \vec{x}) $,
\begin{equation}
	h_{a b}(t, \vec{x})=\sum_A \int_{-\infty}^{\infty} d f \int_{S^2} d \hat{\Omega} h_A(f, \hat{\Omega}) e^{i 2 \pi f(t-\hat{\Omega} \cdot \vec{x})} e_{a b}^A(\hat{\Omega}),
\end{equation}
where $ h_A(f, \hat{k}) $ is the Fourier amplitudes and $e_{ab}^A(\hat{k})$ represents the polarization tensors for GWs. Additionally, we introduce $\hat{\Omega}$ as a unit vector denoting a specific direction on the two-sphere and $ \hat{k}=2 \pi f \hat{\Omega} $. The characteristic strain $ h_c(f) $ for GWs is defined as
\begin{equation}
	\left\langle h_A(f, \hat{k}) h_{A^{\prime}}^*\left(f^{\prime}, \hat{k}^{\prime}\right)\right\rangle  =\frac{1}{16 \pi} \delta\left(f-f^{\prime}\right) \delta_{A A^{\prime}} \delta^2\left(\hat{k}, \hat{k}^{\prime}\right)  h_c(f)^{2} f^{-1},
\end{equation}
where $\delta_{AA'}$ indicates that GWs with different polarization states are uncorrelated. $\delta^2(\hat{k}, \hat{k}')$ is the delta function on the sphere, which indicates that GWs propagating in different directions are uncorrelated. The characteristic strain noise amplitude of a single detector is denoted by $h_n(f)$.

\subsubsection{The strain power spectral density }

The  PSD of  GW, $S_h(f)$, is known to have a relationship with  $h_c(f)$~\cite{Thrane:2013oya}
\begin{equation}
	\sqrt{S_h(f)}=h_c(f) f^{-1 / 2}.
\end{equation}
Similarly, an equivalent expression for $ S_n(f) $ can be established
\begin{equation}
\sqrt{S_n(f)}=h_{\mathrm{n}}(f) f^{-1 / 2}.
\end{equation}

\subsubsection{The energy density}
In cosmology, $\Omega_{\mathrm{gw}}(f)$ is usually used to signify the fractional contribution of the GW energy density~\cite{Allen:1997ad}
\begin{equation}
	\Omega_{\mathrm{gw}}(f)=\frac{1}{\rho_c} \frac{d \rho_{\mathrm{gw}}}{d \ln f},
\end{equation}
where $\rho_c$ is the critical energy density. We have the following conversion formula
\begin{equation}\label{ogw}
	\Omega_{\mathrm{gw}}(f)=\frac{2 \pi^2}{3 H_0^2} f^2 h_c^2(f) =\frac{2 \pi^2  S_h(f)}{3 H_0^2} f^3.
\end{equation}
We have listed all the notations and their dimensions in natural units in Appendix.~\ref{sec_nof}. This helps avoid confusion caused by using multiple notations in this section and allows readers to go back to their definitions as needed.

To quantify the detectability of the GW signal in a given model with a SFOPT, by using Eq.~\eqref{opsnr} and Eq.~\eqref{ogw}, the SNR can be calculated as~\cite{Wang:2020jrd}
\begin{equation}
	\mathrm{SNR}_{\mathrm{gw}}=\sqrt{T_t\int_{f_{\min }}^{f_{\max }} d f\left(\frac{h_{100}^2 \Omega_{\mathrm{gw}}}{h_{100}^2 \Omega_{\mathrm{sens}}}\right)^2}\label{eq_snrdef},
\end{equation}
where $h_{100}^2 \Omega_{\text {sens }}$ corresponds to the expected sensitivity for given detectors~\cite{Thrane:2013oya}. Here $h_{100}$ is defined with the Hubble constant,
\begin{equation}
	H_0=h_{100} \times 100 \mathrm{~km} \mathrm{~s}^{-1} \mathrm{Mpc}^{-1},
\end{equation}
where $h_{100}$ eliminates sensitivity to the observed value of the Hubble constant. There is a current Hubble tension in the measured values of the Hubble constant, with $h_{100}=74.03 \pm 1.42$ from la Supernovae observations~\cite{Riess:2019cxk} and $h_{100}=67.4 \pm 0.5$ from CMB (Planck) measurements~\cite{Planck:2018vyg}. In this work, we choose $h_{100}=0.72$ as a representative value. The quantity frequently depicted on sensitivity curves for energy density is $\Omega_{\mathrm{gw}}(f)h_{100}^{2}$, which is used in the next section when calculating the GW spectrum and the detector sensitivity curve. 

To demonstrate the detectability of GW signals, we calculate their SNR and compare it with the threshold SNR value of the detector. The detector can detect the GW signal if the calculated SNR is larger than the threshold. To better understand the sensitivity of a detector, we need to consider its sensitivity curve.

\section{Phase transition gravitational wave of the DFSZ axion model and its SNR at Cosmic Explorer}\label{sec_gwsnr}
In the previous section, we reviewed the detection principles of the SGWB and introduced the sensitivity curves of GW detectors. This lays the foundation for discussing the GW signals generated by the DFSZ axion model, which will be the focus of this section.  We will calculate the GW signals produced in the DFSZ model and evaluate their SNR and detectability at the future GW detectors CE and ET. In this section, we use the finite-temperature effective field theory~\cite{Coleman:1973jx,Dolan:1973qd} to investigate the phase transition dynamics in DFSZ. At zero temperature, the tree-level potential is given by Eq.~\eqref{Vtree1}.

At the scale of PQ symmetry breaking, only the 
$\sigma$ field acquires a VEV, $\langle \sigma\rangle =v_\sigma/\sqrt{2}$. The singlet can be expanded as 
\begin{equation}
	\sigma = \frac{1}{\sqrt{2}}\left(v_\sigma+\sigma^0 + i\eta_\sigma^0\right).
\end{equation}
The doublets can be expressed as 
\begin{equation}
	H_u = \left(\begin{array}{c} h_u^R+i h_u^I \\ \frac{1}{\sqrt{2}} ( h_u^0 + i\eta_u^0) 
	\end{array}\right),
\end{equation}
\begin{equation}
	H_d =  \left(\begin{array}{c} \frac{1}{\sqrt{2}}(h_d^0 + i\eta_d^0) \\ h_d^R+ih_d^I
	\end{array}\right).
\end{equation} 
The free energy density for the phase transition process can be obtained as the finite-temperature effective potential. 
Generally, the finite-temperature effective potential is composed of three parts
\begin{equation}\label{EFp}
	V_{\text {eff }}(\sigma^0, T) \equiv V_{\text {tree }}(  \sigma^0)+V_{\text {CW}}(  \sigma^0)+V_{\text {T}}( \sigma^0, T),
\end{equation}
where $T$ is temperature and $V_{\text {tree }}$ refers to the tree-level potential. $V_{\text {CW}}(\sigma^0)$ denotes the one-loop corrections at the zero temperature. $V_{\text {T}}(\sigma^0, T)$ represents the thermal correction, including the daisy resummation at finite temperature. The tree-level potential can be simplified as
\begin{equation}
	V_{\text {tree }}(\sigma^0) = -\frac{1}{2}\mu_3^2 (\sigma^0)^2 + \frac{1}{4}\lambda_3 (\sigma^0)^4.
\end{equation}
In the $ \overline{MS} $ scheme~\cite{Coleman:1973jx}, the zero-temperature Coleman-Weinberg (CW) potential can be expressed as follows
\begin{equation}
	V_{\mathrm{CW}}( \sigma^0 )=\frac{1}{64 \pi^2} \sum_S n_S m_S^4(  \sigma^0 )\left[\log \frac{m_S^2(  \sigma^0)}{\mu^2}-\frac{3}{2}\right],
\end{equation}
where $m_S(\sigma^0)$ represents the field-dependent mass and  $n_S$ is the degree of freedom ($d.o.f.$).
$\mu$ represents the renormalization scale, and the constant $3/2$ is for scalars. The thermal correction can be expressed as follows~\cite{Carrington:1991hz}	
\begin{equation}
	V_{\text {T}}( \sigma^0, T)=\sum_S \frac{n_S T^{4}}{2 \pi^{2} }  J_{B}\left[\frac{m_S^{2}( \sigma^0 )}{T^{2}}\right],
\end{equation}
where $ J_{B} $ is the thermal function for bosons
\begin{equation}\label{eq_JBdef}
	J_{B }=\int_0^{\infty} \mathrm{d} x x^2 \log \left[1 - e^{-\sqrt{x^2+m_S^2(\sigma^0) / T^2}}\right].
\end{equation}
In finite-temperature quantum field theory, when the temperature is high, the infrared amplitude of the fields becomes very large. As a result, this leads to the breakdown of perturbative expansion, a problem known as infrared divergence. To solve this problem, we use daisy resummation to improve the effective potential. Daisy resummation resolves the infrared divergence by resumming the self-energy diagrams of the fields. There are usually two schemes for daisy resummation: the Arnold-Espinosa scheme~\cite{Arnold:1992rz} and the Parwani scheme~\cite{Parwani:1991gq}. In this work, we choose the Parwani scheme. In the Parwani scheme, the contribution of daisy resummation to the thermal corrections is represented by $\Pi_S^{ij}$, which can be expressed as~\cite{Basler:2018cwe}
\begin{equation}
	\Pi_{S}^{ij}= \frac{T^{2}}{12}\sum_{k=1}^{n_{\text {scalars }}} L^{i j k k},
\end{equation}
where $n_{\text {scalars }}$ represents the number of scalar fields. Furthermore, the $ L^{i j k k} $ tensor can be defined as follows
\begin{equation}
	\mathcal{L}_S=-L^i \Phi_i-\frac{1}{2 !} L^{i j} \Phi_i \Phi_j-\frac{1}{3 !} L^{i j k} \Phi_i \Phi_j \Phi_k-\frac{1}{4 !} L^{i j k l} \Phi_i \Phi_j \Phi_k \Phi_l.
\end{equation}
The Lagrangian $ \mathcal{L}_S $ represents the model containing scalar fields $ \Phi_i $, with $ i= {h_u^R}, {h_u^0}, {\eta_u^0},  {h_d^I}$, ${h_d^0}, {\eta_d^0}, {\sigma^0},{\eta_\sigma^0},{h_u^I}, {h_d^R} $. The term $ L^{i j k l} $ denotes the coefficient corresponding to the interaction term $ \Phi_i \Phi_j \Phi_k \Phi_l $. The thermal corrections to the scalar fields  are provided below
\begin{align}
	\begin{split}
		&\Pi_{S}^{00}:\frac{20~T^{2}}{12}\lambda_{1}+\frac{12 ~T^{2}}{12}\lambda_{12}+\frac{4 ~T^{2}}{12}\lambda_{13}+\frac{4 ~T^{2}}{12}\lambda_{4},\\
		&\Pi_{S}^{11}:\frac{11 ~T^{2}}{12}\lambda_{1}+\frac{6 ~T^{2}}{12}\lambda_{12}+\frac{2 ~T^{2}}{12}\lambda_{13}+\frac{4 ~T^{2}}{12}\lambda_{4},\\
		&\Pi_{S}^{22}:\frac{11 ~T^{2}}{12}\lambda_{1}+\frac{6 ~T^{2}}{12}\lambda_{12}+\frac{2 ~T^{2}}{12}\lambda_{13}+\frac{4 ~T^{2}}{12}\lambda_{4},\\
		&\Pi_{S}^{33}:\frac{20 ~T^{2}}{12}\lambda_{2}+\frac{12 ~T^{2}}{12}\lambda_{12}+\frac{4 ~T^{2}}{12}\lambda_{23}+\frac{4 ~T^{2}}{12}\lambda_{4},\\
		&\Pi_{S}^{44}:\frac{11 ~T^{2}}{12}\lambda_{2}+\frac{6 ~T^{2}}{12}\lambda_{12}+\frac{2 ~T^{2}}{12}\lambda_{23}+\frac{4 ~T^{2}}{12}\lambda_{4},\\
		&\Pi_{S}^{55}:\frac{11 ~T^{2}}{12}\lambda_{2}+\frac{6 ~T^{2}}{12}\lambda_{12}+\frac{2 ~T^{2}}{12}\lambda_{23}+\frac{4 ~T^{2}}{12}\lambda_{4},\\
		&\Pi_{S}^{66}:\frac{3 ~T^{2}}{12}\lambda_{3}+\frac{6 ~T^{2}}{12}\lambda_{23}+\frac{6 ~T^{2}}{12}\lambda_{13},\\ 
		&\Pi_{S}^{77}:\frac{3 ~T^{2}}{12}\lambda_{3}+\frac{6 ~T^{2}}{12}\lambda_{23}+\frac{6 ~T^{2}}{12}\lambda_{13},\\
		&\Pi_{S}^{88}:\frac{20 ~T^{2}}{12}\lambda_{1}+\frac{12 ~T^{2}}{12}\lambda_{12}+\frac{4 ~T^{2}}{12}\lambda_{13}+\frac{4 ~T^{2}}{12}\lambda_{4},\\
		&\Pi_{S}^{99}:\frac{20 ~T^{2}}{12}\lambda_{2}+\frac{12 ~T^{2}}{12}\lambda_{12}+\frac{4 ~T^{2}}{12}\lambda_{23}+\frac{4 ~T^{2}}{12}\lambda_{4}.\\     
	\end{split}
\end{align}
We calculate the field-dependent mass (see Appendix.~\ref{sec_Fdm}) at finite temperature and subsequently perform the following substitution in the Parwani method~\cite{Parwani:1991gq}
\begin{equation}
	m_S^2 \rightarrow \bar{m}_S^2=m_S^2+\Pi_S^{ij}.
\end{equation}

After obtaining the effective potential, we can use it to calculate some important phase transition parameters, such as the nucleation temperature, the characteristic length scale, the time scale of phase transition, and phase transition strength. The characteristic length is the mean bubble separation $ R $. Typically, we utilize the approximation of the inverse time scale of phase transition $\beta$ to estimate $ R $, given by
\begin{equation}\label{key}
	R=\frac{(8 \pi)^{1 / 3}}{\beta} v_w.
\end{equation}
$ v_w$ is the bubble wall velocity, which is the most crucial parameter in determining the spectrum of phase transition GW, as it directly affects the energy budget~\cite{Espinosa:2010hh,Wang:2023jto}.
Previous studies like~\cite{Moore:1995si,Megevand:2009gh,Huber:2013kj,Konstandin:2014zta,Dorsch:2018pat,Wang:2020zlf,Jiang:2022btc} provide methods for calculating the bubble wall velocity. We will discuss it in detail later.

The inverse time scale of phase transition $\beta$ can be expressed as
\begin{equation}
	\beta=-\left.\frac{d}{d t}\left(\frac{S_3(T)}{T}\right)\right|_{T=T_n}=\left.H(T) T \frac{d}{d T}\left(\frac{S_3}{T}\right)\right|_{T=T_n}.
\end{equation}
$T_n$ is the nucleation temperature. $S_3$ represents the bounce action for a three-dimensional Euclidean space~\cite{Linde:1980tt,Linde1983},
\begin{equation}
	S_3=4 \pi \int_0^{\infty} d r r^2\left[\frac{1}{2}\left(\frac{d \sigma^0}{d r}\right)^2+V_{\mathrm{eff}}(\sigma^0, T)\right].
\end{equation} 
The nucleation rate per unit time and volume, represented by $\Gamma$~\cite{Linde:1980tt}, can be approximated as $\Gamma \approx \Gamma_0 e^{-S_3}$, where $\Gamma_0$ is proportional to $T^4$. 
To calculate the bubble nucleation rate, we need to obtain the bubble profiles of $ v_\sigma $ by solving  the bounce equation with the boundary conditions
\begin{equation}
	\frac{d^2 \sigma^0}{d r^2}+\frac{2}{r} \frac{d \sigma^0}{d r}=\frac{\partial V_{\mathrm{eff}}}{\partial \sigma^0}, \quad \left.\frac{d \sigma^0}{d r}\right|_{r=0}=0, \quad \sigma^0\left|_{r=\infty}=\sigma^0_{\text{false}}\right.,
\end{equation}
where $\sigma^0_{\text{false }}$ is the scalar field value  in the false vacuum. The overshooting/undershooting approach~\cite{Coleman:1977py,Callan:1977pt} is commonly employed to solve the equation for the single-field bounce.  The nucleation temperature $T_n$ corresponds to the temperature at which one bubble is nucleated in one casual Hubble volume during the phase transition. It is defined as the temperature at which the nucleation rate $\Gamma$ is equal to the fourth power of the Hubble rate $H$
\begin{equation}
	\frac{\Gamma\left(T_{n}\right)}{H\left(T_{n}\right)^{4}}=1.
\end{equation}
The Hubble rate $ H(T) $ during the radiation-dominated era is
\begin{equation}
	H(T)=\left[\left(\frac{\pi^{2}}{90}\right) g_{*}\right]^{1 / 2} \frac{T^{2}}{M_{p}},
\end{equation} 
where $  M_{p} $ is reduced Planck mass and $g_{*}$ represents the effective number of relativistic $d.o.f.$. Within the bag model framework, the phase transition strength parameter $ \alpha $ is typically defined as~\cite{Wang:2020jrd} 
\begin{equation}
	\alpha=\left.\frac{\Delta V_{\mathrm{eff}}-T \frac{\partial \Delta V_{\mathrm{eff}}}{\partial T}}{\rho_R}\right|_{T=T_n}.
\end{equation}
To derive $ \alpha $, we can utilize the expressions for the effective potential $V_{\mathrm{eff}}$ and its derivative concerning $T_n$. The effective energy density $\rho_R$ can be expressed as $\rho_R=\frac{\pi^2 g_{*} T_{n}^4}{30}$. $\Delta V_{\mathrm{eff}}$ refers to the difference in effective potential between the symmetric phase and the broken phase at $T_n$. 

The GW spectrum, which is generated by three primary processes: bubble collisions, sound waves in the plasma, and turbulence, can be accurately calculated by using these key phase transition parameters derived from the effective potential.
\begin{itemize}
	\item Bubbles collisions\\
	According to the envelope approximation~\cite{Kosowsky:1991ua, Kosowsky:1992rz,Kosowsky:1992vn}, the shape of the bubble walls is the key factor determining the generation of GWs. Based on the numerical simulation, the GW spectrum from bubble collisions is~\cite{Caprini:2015zlo, Huber:2008hg}
	\begin{equation}
		h_{100}^2 \Omega_{\mathrm{co}}(f) \simeq 1.67 \times 10^{-5}\left(\frac{H_* R}{(8 \pi)^{1 / 3}}\right)^2\left(\frac{\kappa_\sigma \alpha}{1+\alpha}\right)^2\left(\frac{100}{g_*}\right)^{1 / 3} \frac{0.11 v_w}{0.42+v_w^2} \frac{3.8\left(f / f_{\mathrm{co}}\right)^{2.8}}{1+2.8\left(f / f_{\mathrm{co}}\right)^{3.8}},
	\end{equation}
	where $\kappa_\sigma$ represents the fraction of vacuum energy converted into the gradient energy of the scalar field, and $H_*$ is the Hubble parameter at $T_n$. The peak frequency of the GW spectrum generated by bubble collisions is
	\begin{equation}
		f_{\mathrm{co}} \simeq 1.65 \times 10^{-5} \mathrm{~Hz} \frac{(8 \pi)^{1 / 3}}{H_* R}\left(\frac{0.62 v_w}{1.8-0.1 v_w+v_w^2}\right)\left(\frac{T_n}{100~ \mathrm{GeV}}\right)\left(\frac{g_*}{100}\right)^{1 / 6}.
	\end{equation}
	\item Sound wave\\
   The contribution from sound waves arises from the compression and rarefaction of the plasma as the bubble walls expand~\cite{Hindmarsh:2016lnk,Konstandin:2017sat,Hindmarsh:2019phv}. The numerical simulation~\cite{Hindmarsh:2013xza,Hindmarsh:2015qta,Hindmarsh:2017gnf} gives the GW spectrum as
	\begin{equation}\label{eq_swgwder}
		\begin{aligned}
			h_{100}^2 \Omega_{\mathrm{s w}}(f) \simeq & 2.65 \times 10^{-6} {\beta}^{-1}\left(\frac{\kappa_{sw} \alpha}{1+\alpha}\right)^2\left(\frac{100}{g_*}\right)^{1 / 3} v_w\left(f / f_{s w}\right)^3\left(\frac{7}{4+3\left(f / f_{s w}\right)^2}\right)^{7 / 2},
		\end{aligned}
	\end{equation}
	where $\kappa_{sw}$ represents the fraction of vacuum energy that is converted into the kinetic energy of the fluid's bulk motion, and the peak frequency of the GW spectrum generated by sound waves is given by  
	\begin{equation}
		f_{\mathrm{sw}} \simeq 2.6 \times 10^{-5} \mathrm{~Hz} \frac{1}{H_* R}\left(\frac{T_n}{100~ \mathrm{GeV}}\right)\left(\frac{g_*}{100}\right)^{1 / 6}.
	\end{equation}
	\item Turbulence\\
	As the fluid propagates, the complex physical process of turbulence can be triggered. Based on the numerical simulation, the Kolmogorov-type turbulence
	contribution to the GW spectrum is~\cite{Caprini:2009yp,RoperPol:2019wvy}
	\begin{equation}
		h_{100}^2 \Omega_{\mathrm{turb}}(f) \simeq 1.14 \times 10^{-4} H_* R\left(\frac{\kappa_{\mathrm{turb}} \alpha}{1+\alpha}\right)^{3 / 2}\left(\frac{100}{g_*}\right)^{1 / 3} \frac{\left(f / f_{\text {turb }}\right)^3}{\left(1+f / f_{\text {turb }}\right)^{11 / 3}\left(1+8 \pi f / H_*\right)},
	\end{equation}	
	where $ \kappa_{\mathrm{turb} }$ represents how effectively vacuum energy is converted into turbulent flow
	\begin{equation}
		\kappa_{\text {turb }}=\tilde{\epsilon} \kappa_{sw}.
	\end{equation}
	We set $ \tilde{\epsilon} =0.1 $. The peak frequency of the GW spectrum generated by turbulence is given by  
	\begin{equation}
		f_{\text {turb }} \simeq 7.91 \times 10^{-5} \mathrm{~Hz} \frac{1}{H_* R}\left(\frac{T_*}{100~ \mathrm{GeV}}\right)\left(\frac{g_*}{100}\right)^{1 / 6}.
	\end{equation}
\end{itemize}
By calculating the phase transition parameters from the effective potential and combining the contributions from bubble collisions, sound waves, and turbulence, the total GW spectrum is 
\begin{equation}\label{bess}
	h_{100}^2 \Omega_{\mathrm{total}}(f)=h_{100}^2 \Omega_{\mathrm{co}}(f)+  h_{100}^2 \Omega_{\mathrm{s w}}(f)+h_{100}^2 \Omega_{\mathrm{turb}}(f). 
\end{equation}
For the thermal phase transition, the sound wave is the main source of the GW spectra. From Eq.\eqref{eq_swgwder}, we can see that the GW spectra depend on the bubble wall velocity $v_w$, the phase transition strength $\alpha$, the inverse time scale of the phase transition $\beta$, the nucleation temperature $T_n$, and the efficiency parameter $\kappa_{sw}$. $\kappa_{sw}$ is highly sensitive to $v_w$ and can be estimated as following,
\begin{equation}\label{vss}
\kappa_{sw}=\frac{3}{\epsilon v_w^3} \int w(\xi) v_u^2 \gamma_u^2 \xi^2 d \xi,
\end{equation}
where $w(\xi)$ is the enthalpy density,  $v_u$ is the fluid velocity, $\gamma_u \equiv 1/\sqrt{1-v_u^2}$ is the Lorentz factor,  $\epsilon$ is the vacuum energy density, and $ \xi$ is the ratio of the distance from the bubble center to the time since nucleation. Thus, from Eq.\eqref{eq_swgwder} and Eq.\eqref{vss} we can see that $v_w$ has the most significant impact on the GW spectra compared to other parameters. Therefore, uncertainties in the prediction of GW spectrum arise mainly from uncertainties in the prediction of the bubble wall velocity. We  will  
 discuss this in detail in Sec.~\ref{sec_fm}.
\begin{figure}
    \centering
    \includegraphics[width=0.9\linewidth]{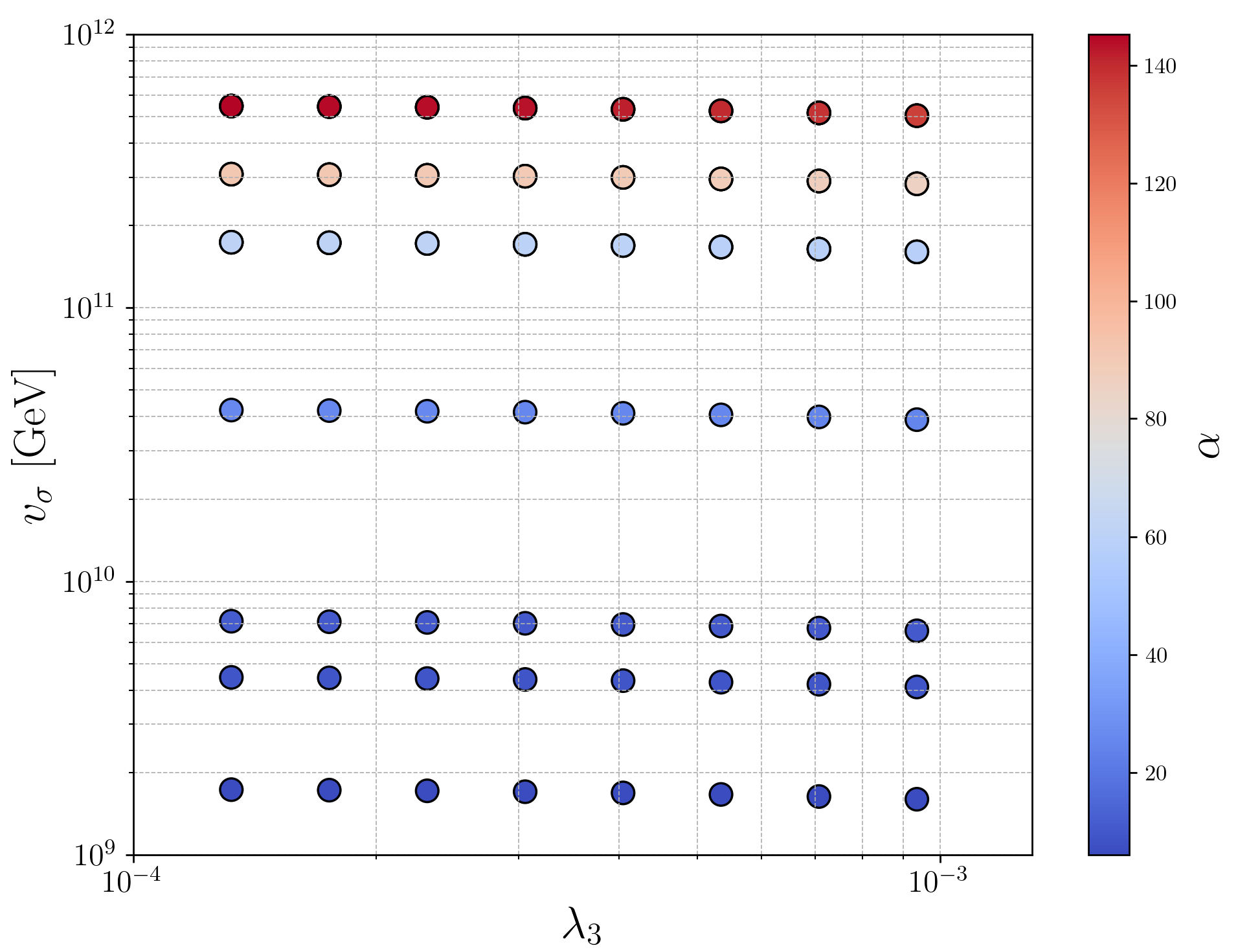}
    \caption{Parameter space exploration results of the SFOPT in the DFSZ axion model. The parameters $\lambda_3$ and $v_\sigma$ are varied over the ranges $10^{-4} \leq \lambda_3 \leq 10^{-3}$ and $10^9~ \mathrm{GeV} \leq v_\sigma \leq$ $10^{12} ~\mathrm{GeV}$, respectively. Each point plotted in the figure represents a set of parameters for which a SFOPT occurs. The color of each point in the figure indicates the value of $\alpha$, with the color bar ranging from blue to red.}
    \label{ES}
\end{figure}
\begin{table}[t]
\begin{center}
    \centering
    \caption{Four sets of benchmark parameters for the DFSZ model.}\label{tab_bp}
    \begin{tabular}{ccccccccc}
        \hline\hline
        & \( \lambda_{3} \) & \( \lambda_{13} \) & \( \lambda_{23} \) & \( \mu \) [GeV] & \( T_{n} \) [GeV] & \( v_{\sigma}(T_{n}) \) [GeV] & \( \alpha \) & \( \beta^* \) \\
        \hline\hline
        \( \text{BP}_1 \) & \( 1.01 \times 10^{-4} \) & \( 0.51 \) & \( 0.14 \) & \( 4.97 \times 10^{17} \) & \( 6.73 \times 10^{7} \) & \( 1.24 \times 10^{9} \) & \( 7.37 \) & \( 48.39 \) \\
        \( \text{BP}_2 \) & \( 2.01 \times 10^{-4} \) & \( 0.61 \) & \( 0.09 \) & \( 1.97 \times 10^{19} \) & \( 5.68 \times 10^{8} \) & \( 7.01 \times 10^{9} \) & \( 1.99 \) & \( 54.60 \) \\
        \( \text{BP}_3 \) & \( 5.01 \times 10^{-4} \) & \( 0.45 \) & \( 0.15 \) & \( 8.48 \times 10^{19} \) & \( 4.18 \times 10^{8} \) & \( 1.62 \times 10^{10} \) & \( 121.29 \) & \( 28.37 \) \\
        \( \text{BP}_4 \) & \( 2.90 \times 10^{-4} \) & \( 0.52 \) & \( 0.15 \) & \( 8.49 \times 10^{21} \) & \( 5.28 \times 10^{9} \) & \( 1.57 \times 10^{11} \) & \( 53.51 \) & \( 28.50 \) \\
        \hline\hline
    \end{tabular}
\end{center}
\end{table}

We perform numerical calculations on the phase transition dynamics using the CosmoTransitions package~\cite{Wainwright:2011kj} to find the parameter space to generate a strong phase transition.  Our goal is to find the SFOPT at high energy scales. During the calculations, we impose the following optimization conditions on the parameters: 1. The critical temperature $T_c$ should be as high as possible and at least greater than the electroweak scale $100~\mathrm{GeV}$. 2. $v_\sigma$ should be significantly greater than $246~\mathrm{GeV}$, as the VEV is related to the energy scale of the phase transition. 3. The ratio of the scalar field mass to the nucleation temperature should be much smaller than 1 to ensure the applicability of Eq.~\eqref{eq_JBdef}. We select parameters based on qualitative analysis. During the phase transition, only the $\sigma$ field acquires a VEV, which is primarily determined by $\mu_3$ and $\lambda_3$ while other quartic coupling terms (such as $\lambda_{12}$, $\lambda_{2}$, etc.) are not sensitive to the VEV of the $\sigma$ field. We find these parameters $\lambda_{1}$, $\lambda_{4}$, $\lambda_{12}$, $\lambda_{5}$, and $\lambda_{2}$  have a smaller impact on the resulting phase transition GW signals in calculations.  Thus the different values between $\mathcal{O}(10^{-1})$ and $\mathcal{O}(10^{-4})$ for these couplings do not significantly affect the results of the phase transition. For simplicity, we set the following values 
\begin{equation}
\lambda_{1}=\lambda_{4}=\lambda_{12}=\lambda_{5}=\lambda_{2}=0.1,
\end{equation}
\begin{equation}
	\mu_{1}=\mu_{2}= 1.0 \times10^{5}~\mathrm{GeV},
\end{equation}
\begin{equation}
\mu_3 =  1.19 \times10^{6}~\mathrm{GeV}.
\end{equation}
Other remaining model parameters, including $\lambda_{3}$, $\mu$, $\lambda_{13}$, and $\lambda_{23}$, are varied to generate different phase transition parameters. To get a larger VEV, a smaller $\lambda_3$ is required since $v_\sigma \propto \dfrac{1}{\sqrt{\lambda_3}}$. Moreover, a smaller $\lambda_3$ indicates that the coupling between axions and Standard Model particles is extremely weak, satisfying the stringent constraints from current experiments and astronomical observations on axion. And considering $\lambda_3$ cannot be excessively small, we choose $\lambda_3$ in the range of $\mathcal{O}(10^{-4})$ to $\mathcal{O}(10^{-3})$. Since the coupling $\lambda_{13}$ and $\lambda_{23}$ affect the nature of the phase transition through their interaction with the $\sigma$ field, larger values are required to promote a stronger phase transition. Therefore, we primarily select values in the range of $\mathcal{O}(10^{-2})$ to $\mathcal{O}(10^{-1})$ for these parameters.

We present the main results of our parameter space exploration in Fig.~\ref{ES}. We map the parameter space where  $\lambda_3$ ranges from $10^{-4}$ to $10^{-3}$, and $v_\sigma$ varies from $10^9~\mathrm{GeV}$ to $10^{12}~\mathrm{GeV}$. Each point plotted in the figure represents a set of parameters for which a SFOPT occurs. The color of each point indicates the value of $\alpha$, with the color bar ranging from blue to red. The redder a point appears, the larger the value of $\alpha$, and consequently, the stronger the phase transition. This figure demonstrates that there exists a substantial parameter space for realizing a SFOPT in the DFSZ model. Moreover, in the Appendix.~\ref{sec_eps}, we provide extended parameter space plots for energy scales up to $10^{14}~\mathrm{GeV}$, further illustrating the potential for SFOPTs at even higher energy scales within the DFSZ model. Based on the results presented in Fig.\ref{ES}, we select four representative benchmark parameter sets ($BP_1$, $BP_2$, $BP_3$, and $BP_4$) for further study, as detailed in Tab.\ref{tab_bp}. These sets are selected based on two main criteria: 1. Consistency with the current experimental constraints on the axion-photon coupling, $g_{a \gamma \gamma}$. 2. The ability to generate a SFOPT. Using these parameter sets, we calculate the GW spectra, which are compared to the sensitivity curves of the CE and ET detectors to determine if these detectors can observe the GW signals.

The PQ symmetry-breaking scale, $f_{PQ}$, is a crucial parameter in axion models, as the axion mass is inversely proportional to $f_{PQ}$~\cite{Gorghetto:2018ocs}, the high-energy scale result indicates that it may be possible for invisible axions to exist in the DFSZ model
\begin{table}[t]
	\begin{center}
		\centering 
		\caption{$m_a$,  $g_{a\gamma\gamma}$ for different $f_{PQ}$ values.}
		\label{tab_ma} 
		\begin{tabular}{cccc}
			\hline\hline
			& $ f_{PQ} $~[GeV] & $ m_{a} $~[eV] & $ g_{a \gamma \gamma} $~[GeV$^{-1}$] \\
			
			\hline\hline
			BP\(_1\) & \( 1.24 \times 10^{9} \) & \( 4.59\times 10^{-3} \) & \( 7.01 \times 10^{-13} \) \\
			BP\(_2\) & \(7.01\times 10^{9} \) & \( 8.12\times 10^{-4} \) & \(1.21 \times 10^{-13} \) \\
			BP\(_3\) & \( 1.62 \times 10^{10} \) & \( 3.51 \times 10^{-4} \) & \( 5.37 \times 10^{-14} \) \\
			BP\(_4\) & \(1.57 \times 10^{11} \) & \( 3.63\times 10^{-5} \) & \( 5.54 \times 10^{-15} \) \\
			\hline\hline
		\end{tabular}
	\end{center}
\end{table}
\begin{equation}
	m_a=5.691(51)\left(\frac{10^9~ \mathrm{GeV}}{f_{PQ}}\right)~ \mathrm{meV}.
\end{equation}
The numbers in parentheses represent the uncertainty in the last two digits of the main value.  Given the $m_a$, we can calculate the axion-photon coupling constant $g_{a \gamma \gamma}$. The calculation is as follows~\cite{GrillidiCortona:2015jxo}
\begin{equation}
	g_{a \gamma \gamma}=\frac{\alpha_c}{2 \pi f_{PQ}}\left(\frac{E}{N}-1.92(4)\right)=\left(0.203(3) \frac{E}{N}-0.39(1)\right) \frac{m_a}{\mathrm{GeV}^2},
\end{equation}
where $\alpha_c$ is the fine structure constant, $E/N$ is the ratio of the electromagnetic anomaly coefficient to the color anomaly coefficient. For the  DFSZ model, $E/N$ is $8/3$~\cite{Zhitnitsky:1980tq,Dine:1981rt}.

\begin{figure}[htbp]
	\centering
    \includegraphics[width=1\linewidth]{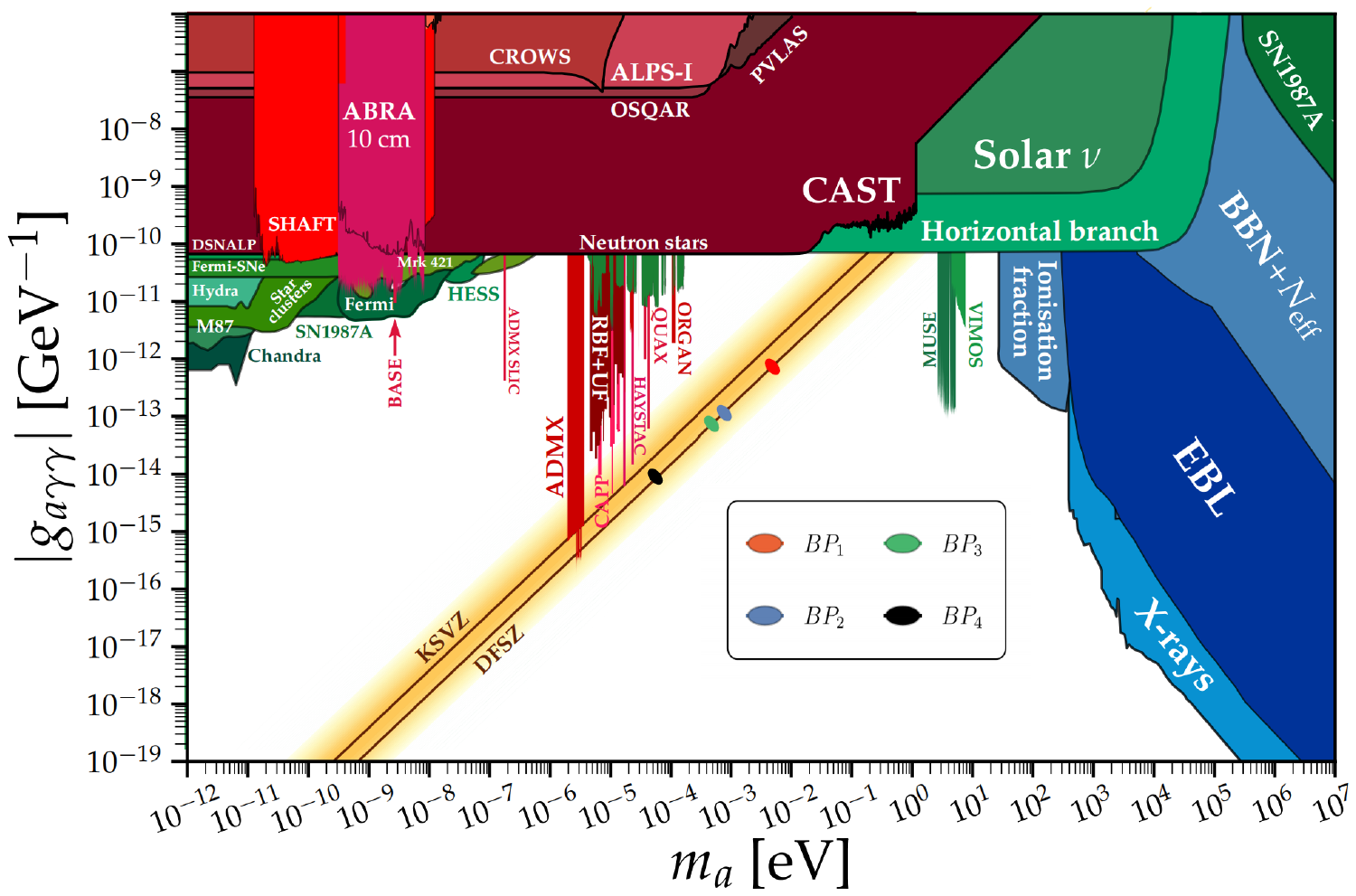}
	\caption{Comparison of the  $m_a$  and the $g_{a\gamma\gamma}$  obtained from all four benchmark points with the current experimental constraints in the ALP-photon coupling exclusion diagram~\cite{OHare2020}.}
	\label{fig:a1}
\end{figure}
Figure~\ref{fig:a1} shows the ALP-photon coupling exclusion plot. The colored regions represent the excluded parameter space from various ALP-photon coupling experiments, such as CAST~\cite{CAST:2017uph}, ADMX~\cite{ADMX:2019uok, ADMX:2020hay}, etc. 
The four elliptical points $BP_{1-4}$ in the plot represent four benchmark parameter sets in Tab.\ref{tab_bp}, shown in red, blue, green, and black, respectively. The corresponding $m_a$ and $g_{a\gamma\gamma}$ values for these points are listed in Tab.~\ref{tab_ma}. The $m_a$ for $BP_1$, $BP_2$, $BP_3$, and $BP_4$ is $4.59\times 10^{-3}~\mathrm{eV}$, $8.12\times 10^{-4}~\mathrm{eV}$, $3.51\times 10^{-4}~\mathrm{eV}$, and $3.63\times 10^{-5}~\mathrm{eV}$, respectively. All four points fall within the expected yellow region, indicating that these points satisfy the constraints from existing experiments. This result identifies parameter sets that are consistent with experimental constraints and the ability to achieve high-energy scale SFOPT.
\begin{figure}[htbp]
	\centering
	\includegraphics[width=0.9\linewidth]{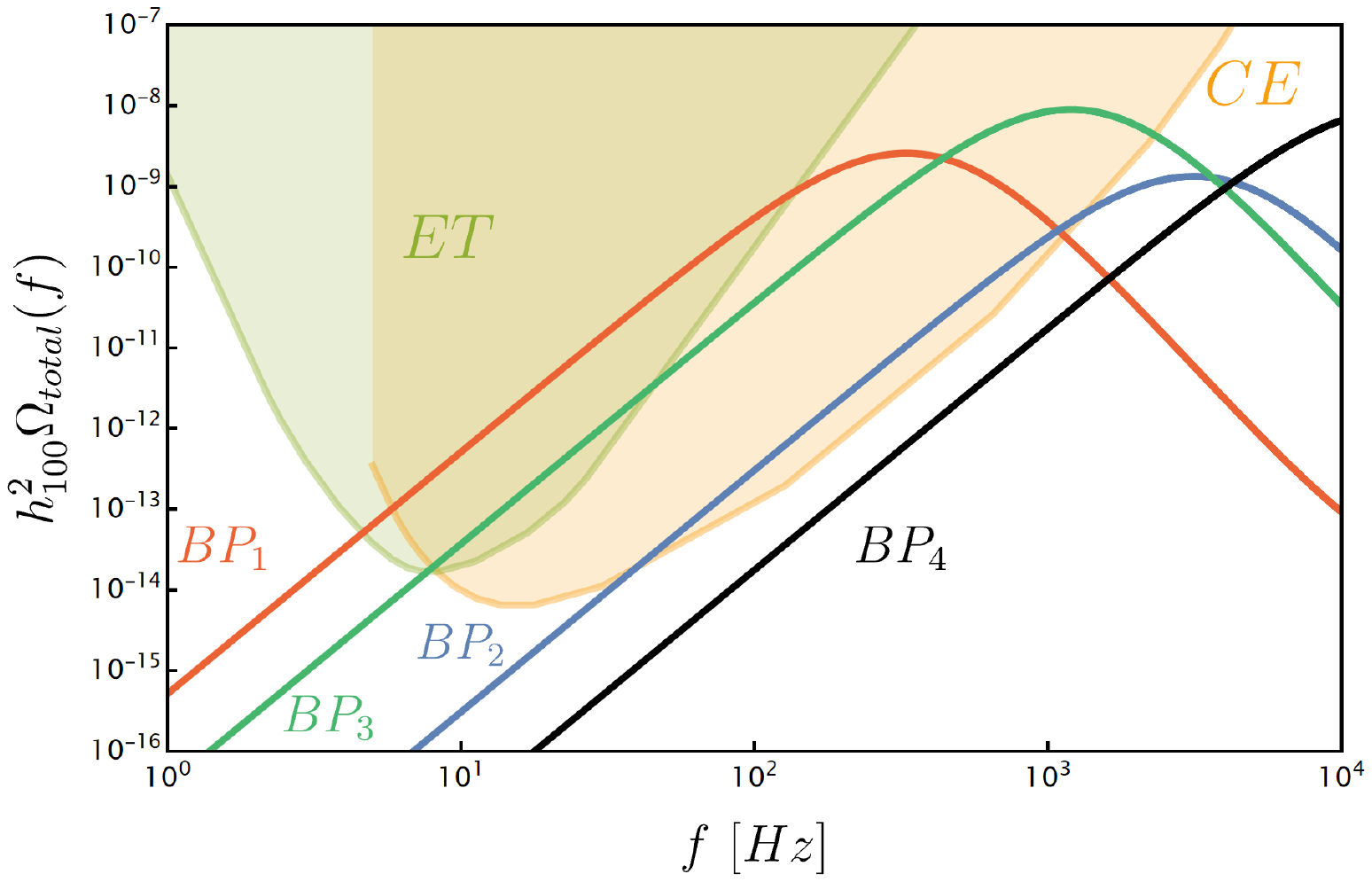} 
	\caption{GW spectra for DFSZ axion model. The light-green and orange curves depict the sensitivity curves with SNR=1 for ET and CE, respectively. The red, blue, green, and black curves represent the phase transition GW signals for $BP_1$, $BP_2$, $BP_3$, and $BP_4$ with $v_w=0.9$. The relevant parameters are summarized in Tab.~\ref{tab_bp}.}
	\label{bp1a}
\end{figure}

Figure~\ref{bp1a} shows the GW spectrum for four sets of benchmark parameters. The light-green and orange curves represent the sensitivity curves for ET and CE with SNR=1 (more details of the sensitivity curves for detectors are introduced in the next section), respectively, over a running time of $1\times10^{8}$ s. The red, blue, green, and black curves represent the GW signals for $BP_1$, $BP_2$, $BP_3$, and $BP_4$ with $v_w=0.9$. Using the definition of Eq.~\eqref{eq_snrdef}, we calculate the SNR values for each benchmark point listed below
\begin{equation}\label{eq_snrval}
	\mathrm{SNR}: \quad BP_1(21794.10)\quad BP_2(46.19)\quad BP_3(3964.97)\quad BP_4(0.15)
\end{equation}
The SNR of $BP_4$ is only 0.15, which is below the threshold value SNR=8~\cite{Hall:2020dps} for the CE detector, making its signal difficult to capture by the CE detector. The SNR of $BP_1$, $BP_2$, and $BP_3$ are 21794.10, 46.19, and 3964.97, respectively, all exceeding the threshold. Thus, the CE detector can detect the signals from these three benchmark points, suggesting that the DFSZ model can generate observable GW signals.
\begin{figure}[htbp]
	\centering
	\includegraphics[width=1\linewidth]{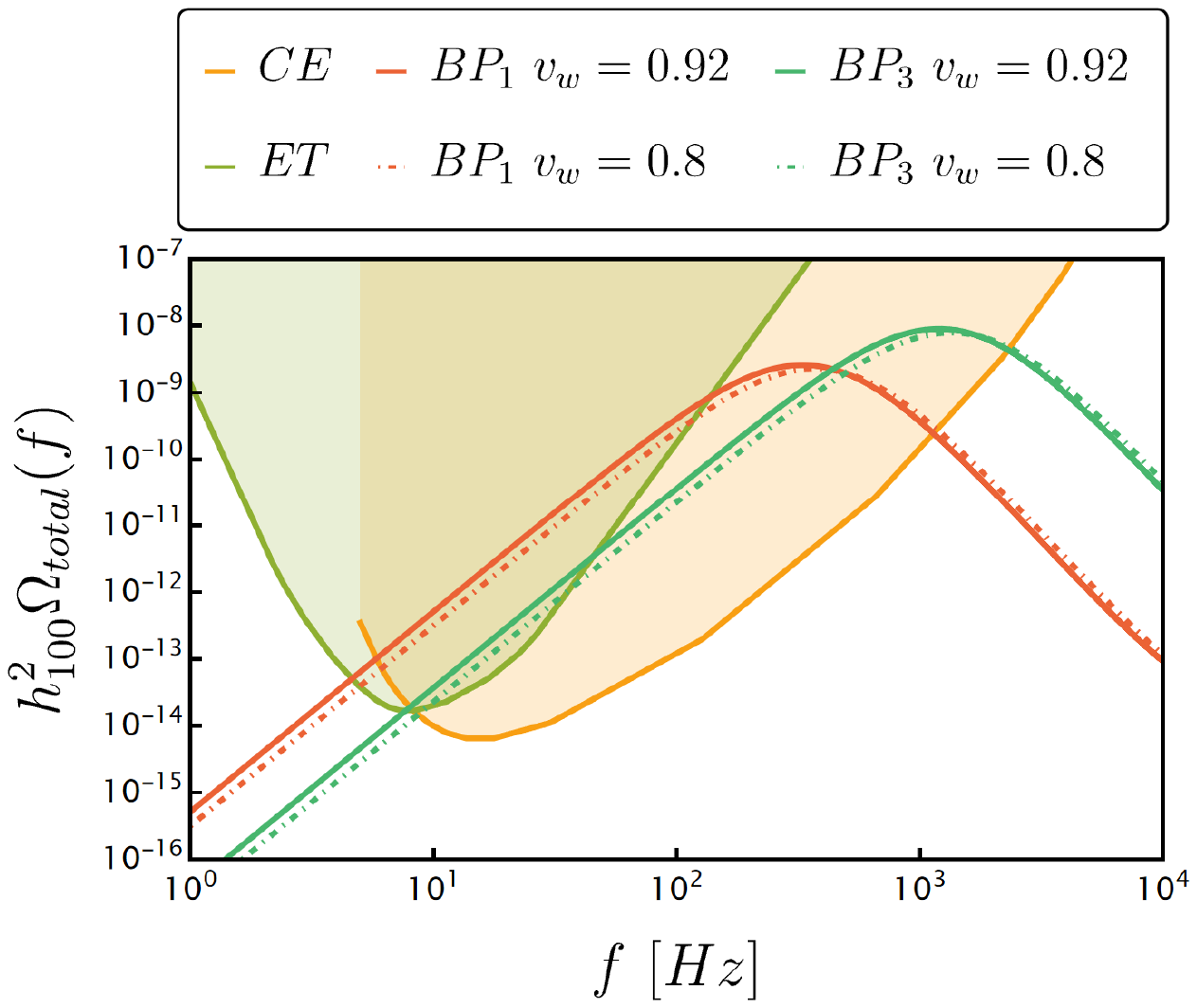} 
	\caption{GW spectra for DFSZ model with $v_w=0.92$ and $v_w=0.8$. The light-green and orange curves depict the sensitivity curves with SNR=1 for ET and CE. The red solid line and the red dash-dotted line correspond to the expected GW signals of $BP_{1}$ with $v_w=0.92$ and $v_w=0.8$, respectively. The green solid line and the green dash-dotted line correspond to the expected GW signals of  $ BP_{3} $ with $v_w=0.92$ and $v_w=0.8$, respectively. The relevant parameters are summarized in Tab.~\ref{tab_bp}.}
	\label{fig_vw}
\end{figure}

The bubble wall velocity has the most significant impact on the spectrum of the phase transition GWs.  Variations in $v_w$ by orders of magnitude can lead to significant changes in the GW spectrum. This makes bubble wall velocity a crucial factor in predicting GW signals. However, the precise calculation of the bubble wall velocity is complicated and difficult. Most previous studies simply take the bubble wall velocity as an input parameter.

Recently, the precise calculation of the bubble wall velocity attracts a lot of attention.  Studies~\cite{Ekstedt:2024fyq,Laurent:2022jrs}  compute the velocity by simultaneously solving the equations of motion of the scalar field and the Boltzmann equations. In particular, Ref.~\cite{Laurent:2022jrs} calculates the bubble wall velocity by using the first principle, and Ref.~\cite{Ekstedt:2024fyq} develops a package based on this method. Ref.~\cite{Dorsch:2023tss} systematically investigates the influence of higher-order terms in the fluid ansatz on bubble wall propagation modes, representing an extension of the classical fluid ansatz. Studies~\cite{Ai:2021kak, Ai:2020seb} propose a model-independent method by deriving upper and lower bounds on the bubble wall velocity under local thermodynamic equilibrium and in the ballistic limit. Additionally, Ref.~\cite{Ai:2023see} confirms that under local thermodynamic equilibrium, the bubble wall velocity is independent of the specific model. Moreover, the study~\cite{BarrosoMancha:2020fay} demonstrates the calculation of forces experienced by expanding bubbles that reach a steady state during SFOPT, derived from first principles of quantum field theory.

Ref.~\cite{Krajewski:2023clt} discovers the phenomenon of hydrodynamical obstruction through lattice simulations in cosmological SFOPT, which excludes certain bubble wall velocities. This indicates that numerical lattice simulations can more accurately determine the possible range of bubble wall velocities.
In further work, authors~\cite{Krajewski:2024gma} investigate the bubble wall velocity under local thermal equilibrium using real-time hydrodynamic simulations. This emphasizes the importance of numerical simulations that consider non-equilibrium dynamics and friction forces for estimating the bubble wall velocity. These findings suggest that simply treating the wall velocity as an input parameter may overlook key dynamical processes such as hydrodynamic obstruction and non-equilibrium effects. Therefore, utilizing numerical simulations and theoretical methods allows for a more accurate estimation of the bubble wall velocity from first principles.

For example, we roughly estimate the bubble wall velocity based on the equilibrium condition~\cite{Marfatia:2020bcs}
\begin{equation}\label{key}
	\Delta V_{\mathrm{eff} }= P_{\mathrm{heavy}}
\end{equation}
where \(\Delta V_{\mathrm{eff}}\) is the potential difference of the phase transition, and \(P_{\mathrm{heavy}}\) is the pressure exerted on the bubble wall where only the contribution from heavy particles is considered,
\begin{equation}
P_{\mathrm{heavy}}=\frac{d_n g_* \pi^2}{90}\left(1+v_w\right)^3 \gamma_\omega^2 T^4.
\end{equation}
Here, \( \gamma_w = \frac{1}{\sqrt{1 - v_w^2}} \) and \( d_n \) is defined by
\begin{equation}
d_n \equiv \frac{1}{g_*}\left[\sum_{ m_i\gtrsim3\gamma_w T}\left(g_i^b+\frac{7}{8} g_i^f\right)\right],
\end{equation}
where the sum accounts for the contributions from particles that are heavy enough.  $g_i^b$ and $g_i^f$ are the $d.o.f.$ of the bosons and fermions, respectively.  For a given $ \alpha $, the equilibrium condition can be rewritten as~\cite{Marfatia:2020bcs}
\begin{equation}\label{key}
	\alpha=\frac{d_n}{3}\left(1+v_w\right)^3 \gamma_\omega^2\,\,.
\end{equation}
We find that for \( \mathrm{BP}_1 \), $v_w = 0.98$, and for \( \mathrm{BP}_2 \), $v_w = 0.94$. This formula actually provides an upper limit on the bubble wall velocity. After considering hydrodynamic obstruction and non-equilibrium effects, the bubble wall velocity will be further reduced. Therefore, selecting $v_w=0.9$ for the GW in Fig.~\ref{bp1a} is reasonable.

To investigate the impact of the unknown parameter $v_w$ on the GW spectrum, we consider two representative values that may produce stronger GW signals during the phase transition process: $v_w=0.92$ and $v_w=0.8$.  Fig.~\ref{fig_vw} shows the GW spectrum for two benchmark parameter sets, $BP_1$ and $BP_3$, that have higher SNR with different $v_w$. The red and green solid (dash-dotted) lines correspond to the expected GW signals of $BP_1$ and $BP_3$ with $v_w=0.92$ ($v_w=0.8$), respectively.

We find differences in the frequency and strength of the GW spectra under different $v_w$ values in Fig.~\ref{fig_vw}. To assess the constraining power of CE on these phase transition parameters more precisely, we need to quantify the uncertainties of each parameter. FM analysis is a powerful statistical tool widely used to estimate the precision with which model parameters can be determined from a given set of observations. In the next section, we will use FM analysis to calculate the relative uncertainties of the phase transition parameters in the DFSZ model and evaluate the constraining power of CE on these parameters.
\begin{table}[t]
	\begin{center}
		\centering 
		\caption{SNR values for two benchmark parameter sets at three different bubble wall velocities~ $v_w$.}\label{tbp4} 
		\begin{tabular}{cccc}
			\hline\hline	
			&$v_w=0.92$ & $v_w=0.9$ & $v_w=0.8$ \\
			\hline\hline
			$ BP_1 $& 23818.07 & 21794.10 & 13486.53\\
			$ BP_3 $& 4375.55 & 3964.97 & 2331.99\\
			\hline\hline
		\end{tabular}
	\end{center}
\end{table}

\section{Fisher matrix analysis of the phase transition parameters at Cosmic Explorer}\label{sec_fm}

FM is a powerful statistical tool widely used for estimating the precision of parameters in various fields, including GW astronomy~\cite{Fisher:1922saa}. For a probability distribution $f(x; \Theta), $ of an observable random variable $x$ that depends on a set of unknown parameters $\Theta = (\Theta_1, \dots, \Theta_k)$, the FM is defined as the expected value of the second-order partial derivatives of the log-likelihood function concerning the parameters. One of the critical properties of the FM is that its inverse corresponds to the covariance matrix of the observable parameters as the probability distribution approximates a Gaussian distribution~\cite{Hamimeche:2008ai}. The diagonal elements of the covariance matrix are the variances (squared uncertainties) of the parameter estimates. Therefore, smaller values of these diagonal elements indicate more precise parameter estimates with lower uncertainties. We will use FM analysis to calculate the relative uncertainties of phase transition parameters, enabling us to assess the constraining power of detectors such as the CE and ET on these parameters. First, we must construct a likelihood model for CE to apply FM analysis.

\subsection{Likelihood model}
In our study, the observational data is the GW signal detected by CE, and the theoretical model parameters are the dynamical parameters of the phase transition generated by the DFSZ axion model. In the following discussion, we will briefly introduce how to construct the likelihood model using CE in preparation for the subsequent FM analysis.

The GW strain in the frequency domain, $\tilde{h}_{gw}\left(f_n\right)$, is given by the discrete Fourier transform of the $h_{gw}(t)$~\cite{Gowling:2021gcy}
\begin{equation}
	\tilde{h}_{gw}\left(f_n\right)=\frac{1}{\sqrt{N_t}} \sum_{m=0}^{N_t-1} h_{gw}(t_m) \exp \left(-2 \pi i f_n t_m\right),
\end{equation}
where $N_t=T_{\mathrm{tot}}/T_{s}$ is the total number of samples, $T_{\mathrm{tot}}=1 \times 10^{8} \mathrm{~s}$ is the total observation time for the CE mission, and $T_{s}$ is the data sampling period. The time $t_m$ is given by $t_m = m T_{s}$. The frequency $f_n$ is defined as
\begin{equation}\label{fn}
	f_n=n / T_{\text {tot }},
\end{equation}
where $n$ is an integer ranging from $-N_t/2$ to $N_t/2$. The GW energy density spectrum, $\Omega_{\mathrm{gw}}\left(f_n\right)$, is related to the strain in the frequency domain by
\begin{equation}
	\Omega_{\mathrm{gw}}\left(f_n\right)=\left(\frac{4 \pi^2}{3 H_0^2}\right) f_n^2\left|\tilde{h}_{gw}\left(f_n\right)\right|^2.
\end{equation}
To optimize computational efficiency, we employ a simplified approach for data grouping known as frequency binning. This involves dividing the frequency range into $N_b$ bins, where the spacing between consecutive bins follows a logarithmic scale. Consequently, each bin contains a specific number of frequencies, denoted as $n_b$ 
\begin{equation}
	n_b=\left[\left(f_b-f_{b-1}\right) T_{\text {tot }}\right],  0 \leq b \leq N_b,
\end{equation}
where $f_b$ and $f_{b-1}$ are the upper and lower frequency boundaries of the $b$-th bin. The PSD of the CE detector can be expressed as $E_n$. The weighted mean value $\bar{E}_b$ of $E_n$ can be expressed as
\begin{equation}
	\bar{E}_b=\frac{A_b}{n_b} \sum_{n \in I_b} \frac{E_n}{A_n},
\end{equation}
where $A_n$ is the variance of the Fourier amplitudes. 
\begin{equation}
	\frac{1}{A_b}=\frac{1}{n_b} \sum_{n \in {I_b}} \frac{1}{A_n},
\end{equation}
$A_b$ is the mean value of the distribution and $I_b$ denote a group of integers $n$ such that $|f_{n}|$ is in frequency bin $b$. For example, considering $I_b=\{1,2,...,m\}$, the variable $\sum_{n \in I_b} E_n/A_n=E_1/A_1+...+E_m/A_m$ approximates a Gamma distribution with shape parameter $2n_b$ when $n_b$ is sufficiently large, according to the Central Limit Theorem. This distribution arises from scaling the Chi-square distribution
\begin{equation}
	p\left(\bar{E}_b \mid A_b\right)=\prod_{b=1}^{N_b} \frac{1}{\left(n_b-1\right) !} \frac{n_b}{A_b}\left(n_b \frac{\bar{E}_b}{A_b}\right)^{n_b-1} \exp \left(-n_b \frac{\bar{E}_b}{A_b}\right).
\end{equation}
Under the Gaussian approximation, the probability distribution is
\begin{equation}
	p\left(\bar{E}_b \mid A_b\right)=\prod_{b=1}^{N_{\mathrm{b}}}\left(\frac{n_b}{2 \pi A_b^2}\right)^{\frac{1}{2}} \exp \left(-\frac{1}{2} \frac{n_b\left(\bar{E}_b-A_b\right)^2}{A_b^2}\right).
\end{equation}
From Eq.~\eqref{eq_logldef} in the Appendix.~\ref{sec_FI}, the natural logarithm of the likelihood function with the Gaussian approximation is
\begin{equation}
	l_G=\ln (p)=-\frac{1}{2} \sum_{b=1}^{N_{\mathrm{b}}} \frac{ n_b\left(\bar{E}_b-A_b\right)^2}{A_b^2}-\sum_{b=1}^{N_{\mathrm{b}}} \ln2 A_b+\text { const. }.
\end{equation}
The FM can be obtained from Eq.~\eqref{eq_fmdef1} is 
\begin{equation}\label{eq_fmdef2}
	\begin{aligned}
		F_{i j}^G= & =E\left[\frac{\partial l_G}{\partial \Theta_i} \frac{\partial l_G}{\partial \Theta_j}\right] \\
		& =E\left[\frac{\partial l_G}{\partial A_b} \frac{\partial l_G}{\partial A_b} \frac{\partial A_b}{\partial \Theta_i} \frac{\partial A_b}{\partial \Theta_j}\right] \\
		& =E\left[\frac{\partial^2}{\partial^2 A_b}\left(-\frac{1}{2} \sum_{b=1}^{N_b} \frac{n_b\left(\bar{E}_b - A_b\right)}{A_b^2}-\sum_{n=1}^{N_b} \ln 2A_b\right)\right]\frac{\partial A_b}{\partial \Theta_i} \frac{\partial A_b}{\partial \Theta_j} \\
		&=\sum_{b=1}^{N_{\mathrm{b}}} \frac{ n_b}{A_b^2} \frac{\partial A_b}{\partial \Theta_i} \frac{\partial A_b}{\partial \Theta_j}\\
		&=T_{\text {obs }} \sum_{b=1}^{N_{\mathrm{b}}} \frac{ \Delta f_b}{\Omega_{\mathrm{t}}^2} \frac{\partial \Omega_{\mathrm{t}}}{\partial \Theta_i} \frac{\partial \Omega_{\mathrm{t}}}{\partial \Theta_j}.
	\end{aligned} 
\end{equation}
To establish the theoretical model for the data, we employ the total power spectrum $\Omega_{\mathrm{t}}$ (defined in Eq.~\eqref{eq_tpsdef}) instead of the spectral densities $A_b$. The covariance matrix is the inverse of Eq.~\eqref{eq_fmdef2}, and the square roots of the diagonal entries correspond to the relative uncertainties of the parameters.

\subsection{ Cosmic Explorer noise}
The CE~\cite{LIGOScientific:2016wof,Reitze:2019iox,Hall:2020dps, Hall:2022dik} and the ET~\cite{ Sathyaprakash:2012jk, Maggiore:2019uih} are third-generation ground-based GW detectors anticipated to commence operations in the 2030s. Compared to second-generation detectors, their sensitivities are expected to increase by one order of magnitude, enabling the detection of frequencies as low as several hertz. This sensitivity enhancement will allow for the observation of $\mathcal{O}(10^5\text{-}10^6)$ compact binary coalescence events (CBCs) within a year-long observation period, including but not limited to binary black hole (BBH) and binary neutron star (BNS) mergers~\cite{LIGOScientific:2016wof}.

As shown in Fig.~\ref{bp1a}, CE and ET can detect GW signals from a SFOPT in axion models. However, we choose CE for the FM analysis, as its sensitivity is higher than that of ET. To better understand the noise model of CE, we need to analyze various noise sources and their impact on the detector's sensitivity.

\subsubsection{shot noise}
The first noise source we consider is the shot noise of the laser, stemming from the quantum nature of photons. In GW detectors, shot noise limits the sensitivity at high frequencies. One method to mitigate the impact of shot noise on experimental results is increasing the number of photons received by the detector, achievable by amplifying the power of the beam, $P_{\text{bs}}$.
The initial beam power, $P_{0}$, refers to the power of the original beam from the laser source incident on the interferometer. A beam splitter is an optical component that divides an incoming beam into two parts, and $C$ represents the multiplier by which the beam power is increased before reaching the beam splitter. Therefore, $P_{\text{bs}} \equiv C P_0 = 1650~\mathrm{kW}$ denotes the power measured at the beam splitter after power amplification.

The shot noise is formulated as~\cite{Maggiore:2007ulw}
\begin{equation}
	\left.S_n^{1 / 2}(f)\right|_{\text {shot }}=\frac{1}{8 \mathcal{F} L}\left(\frac{4 \pi  \lambda_{\mathrm{L}} }{\eta P_{\mathrm{bs}}}\right)^{1 / 2} \sqrt{1+\left(f / f_p\right)^2},
\end{equation}
where $\mathcal{F}$ denotes the Finesse of the Fabry-Perot cavities, a measure of the optical quality and reflectivity of the mirrors within the Fabry-Perot interferometer, set at $\mathcal{F}=200$. $ L $ is arm lengths of $ 40~\mathrm{km} $ and $ \eta=0.96 $ is the efficiency of the photodetector. The pole frequency, $f_{p}=500~\mathrm{Hz}$, is a crucial parameter describing the frequency response of a GW detector, especially those utilizing Fabry-Perot cavities. The symbol $\lambda_{\mathrm{L}}$ denotes the laser wavelength used in the interferometer, which is approximately $1550~\mathrm{nm}$.

\subsubsection{radiation pressure}

In GW detection, enhancing the experimental signal through increased laser beam power is a common practice~\cite{Caves:1980pha}. However, while this approach improves detector sensitivity, it also introduces additional noise in the form of radiation pressure. This pressure is the movement of mirrors as the beam traverses its path and reflects, exerting a force on the mirrors in the system. The strain sensitivity due to radiation pressure typically takes the following form~\cite{Maggiore:2007ulw}
\begin{equation}
	\left.S_n^{1 / 2}(f)\right|_{\mathrm{rad}}=\frac{16 \sqrt{2} \mathcal{F}}{M L(2 \pi f)^2} \sqrt{\frac{1}{2 \pi} \frac{P_{\mathrm{bs}}}{\lambda_L }} \frac{1}{\sqrt{1+\left(f / f_p\right)^2}},
\end{equation}
where $M=320~\text{kg}$ represents the mass of the mirror.

Combining the shot noise and radiation pressure, one obtains the optical readout noise with its PSD given by
\begin{equation}
	\left.S_n(f)\right|_{\text {opt }}=\left.S_n(f)\right|_{\text {shot }}+\left.S_n(f)\right|_{\text {rad}}. 
\end{equation}
So that the energy density is 
\begin{equation}
	\Omega_{\mathrm{opt}}=\left(\frac{4 \pi^2}{3 H_0^2}\right) f^3 S_n(f)|_{\text {opt}}.
\end{equation}

\subsubsection{Seismic noise}

Seismic noise arises from the continuous movement of the ground, with amplitudes around several micrometers. The strain sensitivity due to seismic noise typically can be represented as~\cite{Maggiore:2007ulw}
\begin{equation}
	\frac{x(f)}{L} \simeq \frac{A}{L}\left(\frac{1 \mathrm{~Hz}}{f}\right)^\nu \mathrm{~m} \cdot \mathrm{Hz}^{-1 / 2}.
\end{equation}
Above $1 \mathrm{~Hz}$, the exponent $\nu \simeq 2$, with the magnitude of $A$ around $10^{-7}$. We obtain the noise strain sensitivity by dividing $x(f)$ by the length $L$.
The energy density is 
\begin{equation}
	\Omega_{\mathrm{se}}=\left(\frac{4 \pi^2}{3 H_0^2}\right) f^3 \left(\frac{x(f)}{L}\right)^2.
\end{equation}
Seismic noise predominantly governs the lower frequency domain, while at higher frequencies, shot noise and radiation pressure noise become increasingly significant. The total power spectrum energy density of the CE noise model is 
\begin{equation}
	\Omega_{\mathrm{n}}=\Omega_{\mathrm{opt}}+\Omega_{\mathrm{se}}.
\end{equation}
\subsubsection{Compact binary coalescence}

The GW power spectrum of CBCs foreground, such as BBH and BNS~\cite{LIGOScientific:2017zlf, Rosado:2011kv, Sachdev:2020bkk} is often defined as
\begin{equation}
	\Omega_{\mathrm{cbc}}(f)=\Omega_{\mathrm{ref}}\left(\frac{f}{f_{\mathrm{ref}}}\right)^{2 / 3},
\end{equation}
where $\Omega_{\mathrm{ref}}$ represents the energy spectral density at a reference frequency $f_{\mathrm{ref}}$. Due to the uncertain spatial distribution of BBH and BNS sources, $\Omega_{\mathrm{ref}}$ is uncertain, typically ranging from $10^{-10}$ to $10^{-7}$~\cite{Sachdev:2020bkk}. For this analysis, we assume $\Omega_{\mathrm{ref}} = 10^{-10}$. For BBH mergers, which typically occur within a frequency range from a few hertz to several hundred hertz, we set a reasonable reference frequency $f_{\mathrm{ref}}=200~\mathrm{Hz}$. This frequency indicates the coalescence phase during the BBH merger process, where the GW signal is strongest. For BNS mergers, which happen at slightly higher frequencies than BBH mergers, a higher reference frequency of $ 1000~\mathrm{Hz} $ is selected. 

The total energy density is 
\begin{equation}\label{eq_tpsdef}
	\Omega_{\mathrm{t}}=\Omega_{\mathrm{n}}+\Omega_{\mathrm{cbc}}+\Omega_{\mathrm{gw}}.
\end{equation}

\subsection{The relative uncertainties}
After obtaining Eq.~\eqref{eq_tpsdef} for the CE detector, we can substitute it into Eq.~\eqref{eq_fmdef2} to calculate the relative uncertainty of specific model parameters. We choose the sound wave mechanism, which dominates the GW spectrums. The considered parameters include the bubble wall velocity, the phase transition strength, the Hubble-scaled mean bubble separation $R_H=H_*R$, the nucleation temperature, and the bubble wall velocity. These parameters can be represented in a vector $\vec{\Theta}$, defined as $\vec{\Theta}=\left(v_{\mathrm{w}}, \alpha,  R_H,  \left(T_{\mathrm{n}} /~\mathrm{GeV}\right)\right)$. 
To calculate the relative uncertainties of the GW spectrum generated by the sound wave mechanism, we rewrite Eq.~\eqref{eq_swgwder} in the following form to include the $R_H$,
\begin{equation}
	h_{100}^2 \Omega_{s w}(f) \cong 1.64 \times 10^{-6}\frac{ R_H^2}{\bar{U}_f}
	\left(\frac{\kappa_v \alpha}{1+\alpha}\right)^2\left(\frac{100}{g_*}\right)^{1 / 3}\left(\frac{f}{f_{s w}}\right)^3\left(\frac{7}{4+3\left(f / f_{s w}\right)^2}\right)^{7 / 2},
\end{equation}
with
\begin{equation}
	f_{s w} \cong 2.6 \times 10^{-5} \mathrm{~Hz} \frac{1}{ R_H}\left(\frac{T_n}{100~ \mathrm{GeV}}\right)\left(\frac{g_*}{100}\right)^{1 / 6},
\end{equation}
where $\bar{U}_f^2$ is the root-mean-square fluid velocity~\cite{Caprini:2019egz}
\begin{equation}
	\bar{U}_f^2 \approx \frac{3}{4} \frac{\kappa_v \alpha}{1+\alpha}.
\end{equation}

\begin{figure}[htbp]
	\centering
	\includegraphics[width=0.9\linewidth]{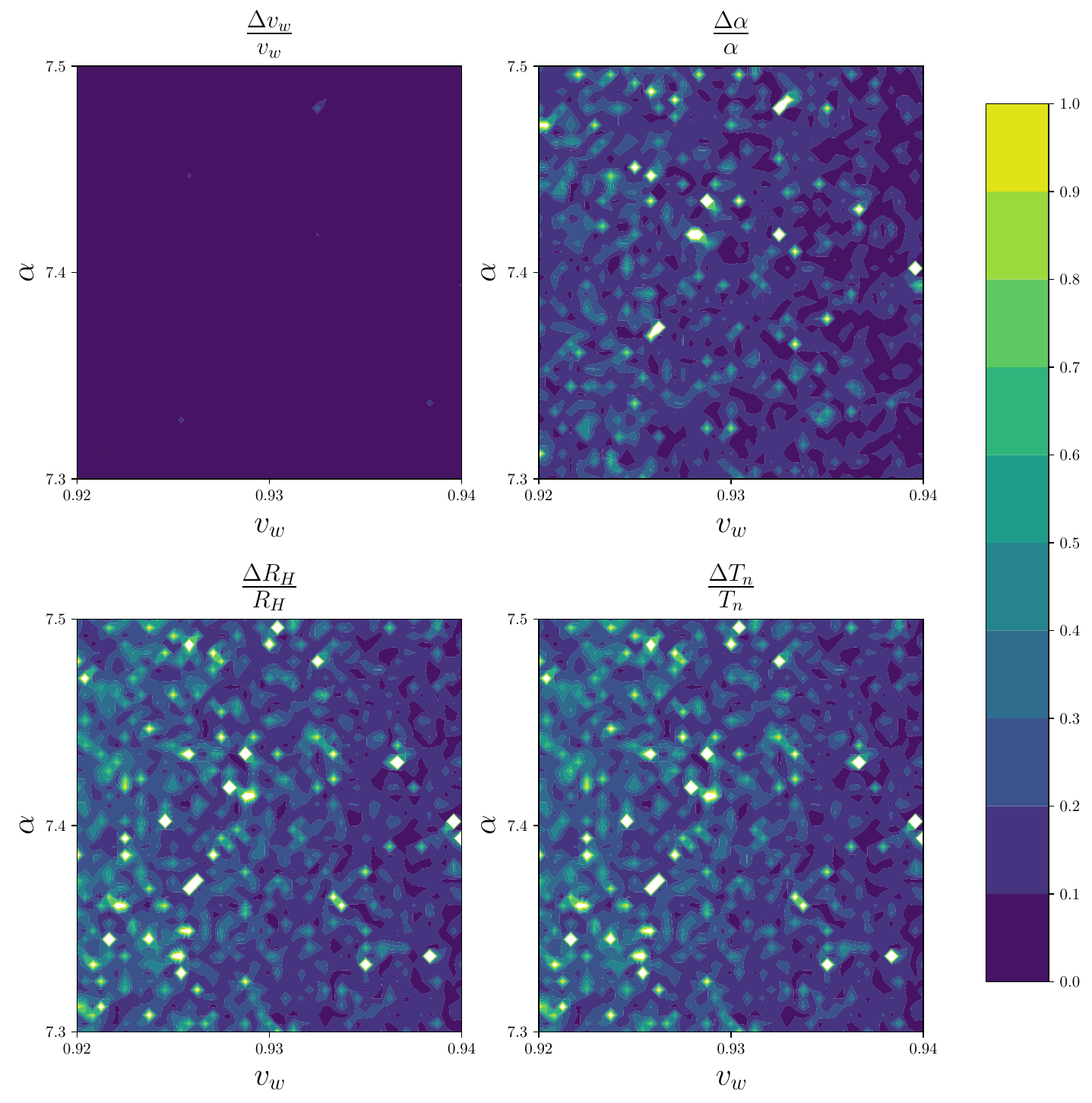}
	\caption{Relative uncertainty with phase transition dynamics parameters the bubble wall velocity $ v_w $, the phase transition strength $ \alpha $, the Hubble-scaled mean bubble spacing $R_H $, and the nucleation temperature $ T_n $. The uncertainty of $v_w$ is roughly an order of magnitude smaller than that of the other parameters. It is obvious that $ v_w $ has a significant impact on the GW spectrum. }\label{ru}	
\end{figure}



We select a set of phase transition parameters that can be detected by CE around $BP_1$ from Tab.\ref{tab_bp}. We set $T_n$ to be
$6.73\times 10^{7}~\mathrm{GeV}$ and
$R_H=0.01$. The parameter space is sampled by varying $v_w$ from 0.92 to 0.94 and $\alpha$ from 7.3 to 7.5.
The relative uncertainties of $v_w$, $\alpha$, $R_H$, and $T_n$ within the given ranges of $v_w$ and $\alpha$  are shown in the four subplots of Fig.~\ref{ru} denoted as $\Delta v_w / v_w$, $\Delta \alpha/\alpha$, $\Delta R_H/R_H$ and $\Delta T_n/T_n$, respectively. A color bar on the right side of the figure indicates the magnitude of the relative uncertainties. From dark blue to light green, the relative uncertainties gradually increase. It can be seen from the figure that the relative uncertainty of $v_w$ is the smallest, as the color in the $v_w$ subplot is the most blue-shifted.

The above FM analysis quantifies how different parameters affect the GW spectrum. When a signal is detected, smaller relative uncertainties in parameter estimation indicate the greater impact on the signal. As shown in Fig.~\ref{ru}, we find that $v_w$ has the smallest relative uncertainties, indicating it has the most significant impact on the GW spectrum. Numerically, we find that the effect of $v_w$ on the GW spectrum is approximately one order of magnitude larger than that of other parameters such as $\alpha$.  As shown in Tab.~\ref{tbp4}, when $v_w$ changes from 0.8 to 0.92, the SNR of the GW signal has obvious variations, and Fig.~\ref{ru} further quantifies this impact. The results obtained through the FM analysis are consistent with those presented in Fig.~\ref{fig_vw}. The numerical results are consistent with the analytical discussions in the paragraph below Eq.\eqref{bess}.

\section{Conclusion}\label{sec_sum}

In this work, we have comprehensively investigated the GW detection of the DFSZ axion model at CE, providing a new way to explore the axion properties. Taking the popular DFSZ axion model as a typical example, we precisely calculate the phase transition dynamics of the $U(1)$ PQ symmetry-breaking process and its associated phase transition GW. 

After calculating the accurate finite-temperature effective potential, we obtain the precise
phase-transition parameters in the DFSZ axion model.
Then, we scan the parameter space in the DFSZ model and find that SFOPT can occur over a broad energy scale
ranging from $10^{9}~\mathrm{GeV}$ to $10^{14}~\mathrm{GeV}$ or even a higher scale. 
We focus on the energy scale from $10^{9}~\mathrm{GeV}$ to $10^{12}~\mathrm{GeV}$ since the peak frequency of the corresponding phase transition spectrum is within the frequency band of the ground-based GW detectors, such as CE and ET.
We calculate the axion mass and axion-photon coupling constant predicted by the DFSZ model and find that they are consistent with current experimental constraints. Then, we quantify the detectability of the DFSZ model at the CE and ET by precisely calculating the SNR of the GW signals. For $v_w=0.9$, signals achieve $\mathrm{SNR}>8$ at CE, indicating that CE can observe them.  

Considering the uncertainty of GW spectra originating from the bubble wall velocity, we provide  GW spectra with different bubble wall velocities as examples.
As shown in Fig.~\ref{fig_vw}, the results show that the GW signals are particularly sensitive to the value of the bubble wall velocity $v_w$. To quantify the uncertainties of phase transition parameters for future GW detectors, we perform a FM analysis over the phase transition parameters $v_w$, $\alpha$, $R_H$, and $T_n$. The results show that the relative uncertainty of $v_w$ is smaller than other parameters, as shown in Fig.~\ref{fig_vw}. Thus, if the predicted phase transition GW spectra are observed at detectors like CE, bubble wall velocity $v_w$ will be the most strongly constrained parameter to be measured.


In conclusion, our findings highlight the substantial potential of GW detection in probing the DFSZ axion model and demonstrate the sensitivity of detectors such as CE to phase transition GWs.

Looking ahead, given that the peak frequency of GWs produced by a SFOPT scales with the energy scale of the phase transition, future detectors with enhanced high-frequency sensitivity could significantly extend the reach of our investigations. Such advancements would enable the detection of higher PQ symmetry-breaking scales and smaller axion masses. These developments promise to deepen our understanding of axion physics and shed light on the evolution of the early universe.

\acknowledgments
We thank Seokhoon Yun for the helpful discussions on axion constraints. A.Y. thanks Siyu Jiang, Jing Yang, Ning Xie, and Chikako Idegawa for their useful suggestions on this manuscript. This work is supported by the National Natural Science Foundation of China (NNSFC) under Grant No.12205387 and No.12475111. 
This work is partly supported by Guangdong Major Project of Basic and Applied Basic Research (Grant No. 2019B030302001).

\appendix 
\section{The field-dependent mass}\label{sec_Fdm}
To determine the field-dependent mass of each physical particle, we need to diagonalize the mass-squared matrix.  The elements of the  mass-squared matrix $M$ for the bosons can be expressed as follows
\begin{align*}
	M_{00} &= v_\sigma^2 \lambda_{13} - 2 \mu_1^2, & M_{11} &= \frac{v_\sigma^2 \lambda_{13}}{2} - \mu_1^2, & M_{22} &= \frac{v_\sigma^2 \lambda_{13}}{2} - \mu_1^2, \\
	M_{33} &= v_\sigma^2 \lambda_{23} - 2 \mu_2^2, & M_{44} &= \frac{v_\sigma^2 \lambda_{23}}{2} - \mu_2^2, & M_{55} &= \frac{v_\sigma^2 \lambda_{23}}{2} - \mu_2^2, \\
	M_{66} &= 3 v_\sigma^2 \lambda_3 - \mu_3^2, & M_{77} &= v_\sigma^2 \lambda_3 - \mu_3^2, & M_{88} &= v_\sigma^2 \lambda_{13} - 2 \mu_1^2, \\
	M_{99} &= v_\sigma^2 \lambda_{23} - 2 \mu_2^2, & M_{03} &= -v_\sigma^2 \lambda_5, & M_{30} &= -v_\sigma^2 \lambda_5, \\
	M_{14} &= \frac{v_\sigma^2 \lambda_5}{2}, & M_{41} &= \frac{v_\sigma^2 \lambda_5}{2}, & M_{25} &= -\frac{v_\sigma^2 \lambda_5}{2},\\
	M_{52} &= -\frac{v_\sigma^2 \lambda_5}{2}, & M_{89} &= \frac{v_\sigma^2 \lambda_5}{2}, & M_{98} &= \frac{v_\sigma^2 \lambda_5}{2}.
\end{align*}
Let
\begin{equation}\label{key1}
	G_{44}= M_{44} +\Pi_{S}^{44},\quad G_{55}= M_{55} +\Pi_{S}^{55},\quad
	G_{66}=M_{66} +\Pi_{S}^{66} .
\end{equation}
\begin{equation}\label{key2}
	G_{77}= M_{77} +\Pi_{S}^{77},\quad G_{88}= M_{88} +\Pi_{S}^{88},\quad
	G_{99}=M_{99} +\Pi_{S}^{99} .
\end{equation}
The mass square matrix $M$ with the daisy corrections is
\begin{equation}
	M=\left(\begin{array}{cccccccccc}
		M_{00} +\Pi_{S}^{00} & 0 & 0 & M_{03} & 0 & 0 & 0 & 0 & 0 & 0\\
		0 & M_{11} +\Pi_{S}^{11} & 0 & 0 & M_{14} & 0 & 0 & 0 & 0& 0\\
		0 & 0 & M_{22} +\Pi_{S}^{22} & 0 & 0 & M_{25} & 0 & 0 & 0 & 0\\
		M_{30} & 0 & 0 & M_{33} +\Pi_{S}^{33} & 0 & 0 & 0 & 0 & 0 & 0 \\
		0 &	M_{41} & 0 & 0 & G_{44} & 0 & 0 & 0 & 0 & 0 \\
		0 &	0 &	M_{52} & 0 & 0 & G_{55} & 0 & 0 & 0 & 0  \\
		0 &	0 &	0 & 0 & 0 & 0 & G_{66}  & 0 & 0 & 0 \\
		0 &	0 &	0 &	0 & 0 & 0 & 0 & G_{77} & 0 & 0  \\
		0 &	0 &	0 &	0 &	0 & 0 & 0 & 0 & G_{88} & M_{89}  \\
		0 &	0 &	0 &	0 &	0 & 0 & 0 & 0 & M_{98} & G_{99}  
	\end{array}\right).
\end{equation}

\section{Exploration of the Extended Parameter Space}\label{sec_eps}
\begin{figure}[htbp]
    \centering
    \includegraphics[width=0.8\linewidth]{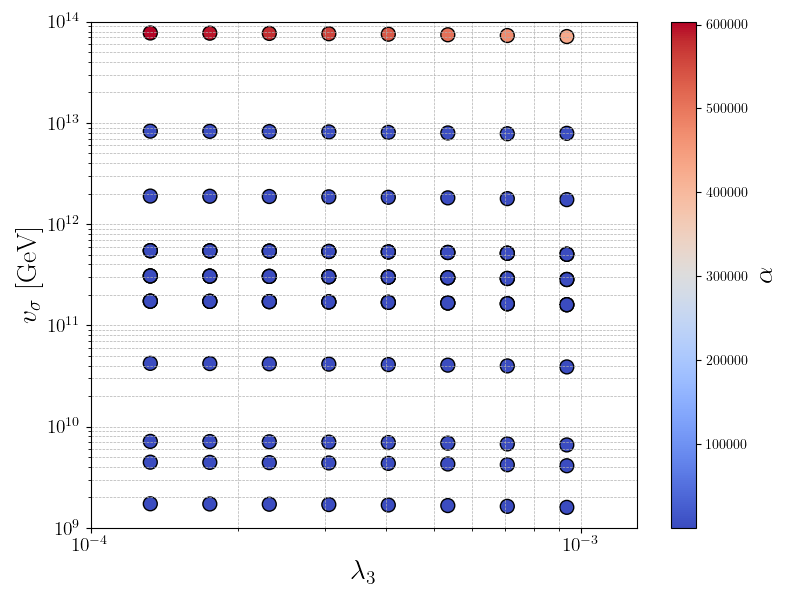}
    \caption{Parameter space exploration of the SFOPT in the DFSZ axion model, without consideration of current experimental constraints on axions. The parameters $\lambda_3$ and $v_\sigma$ are varied over the ranges $10^{-4} \leq \lambda_3 \leq 10^{-3}$ and $10^9~\mathrm{GeV} \leq v_\sigma \leq$ $10^{14} ~\mathrm{GeV}$, respectively.}
    \label{fig_14}
\end{figure}

In the absence of current experimental constraints on axions, we investigate a parameter space where $\lambda_3$ varies between $10^{-4}$ and $10^{-3}$, and $v_\sigma$ varies between $10^9~\mathrm{GeV}$ and $10^{14}~\mathrm{GeV}$. Fig~\ref{fig_14}  demonstrates that there exists a substantial parameter space within the DFSZ model that allows for a SFOPT at high energy scales.

\section{Fisher information}\label{sec_FI}

Once the probability density distribution function $f(\Theta; x)$ is known, the logarithm of the likelihood function  can be expressed as
\begin{equation}\label{eq_logldef}
	l(\Theta ; x)=\ln [f(\Theta ; x)].
\end{equation}
The score function is defined as the partial derivative of the natural logarithm of the likelihood function 
\begin{equation}
	\frac{\partial}{\partial \Theta}l(\Theta ; x).
\end{equation}
Fisher information (FI) represents the second cross-moment of the score vector
\begin{equation}
	F(\Theta)=\mathrm{E}\left[\left(\frac{\partial}{\partial \Theta} l(\Theta ; x)\right) \left(\frac{\partial}{\partial \Theta} l(\Theta ; x)\right)^T\right].
\end{equation}
Given mild regularity conditions, the score function's expected value equals zero
\begin{equation}
	\mathrm{E}\left[\frac{\partial}{\partial \Theta} l(\Theta ; x)\right]=0.
\end{equation}
As a consequence, the FI is defined as the variance of the score 
\begin{equation}
	\begin{aligned}
		F(\Theta) & =\mathrm{E}\left[\left(\frac{\partial}{\partial \Theta} l(\Theta ; x)\right) \left(\frac{\partial}{\partial \Theta} l(\Theta ; x)\right)^T\right]\\
		& =\mathrm{E}\left[\left\{\frac{\partial}{\partial \Theta} l(\Theta ; x)-\mathrm{E}\left[\frac{\partial}{\partial \Theta} l(\Theta ; x)\right]\right\}\left\{\frac{\partial}{\partial \Theta} l(\Theta ; x)-\mathrm{E}\left[\frac{\partial}{\partial \Theta} l(\Theta ; x)\right]\right\}\right] \\
		& =\operatorname{Var}\left[\frac{\partial}{\partial \Theta} l(\Theta ; x)\right].
	\end{aligned}
\end{equation}
It is assumed that the observations in the sample are independent and identically distributed. The maximum likelihood estimator, represented by $\hat{\Theta}$, is approximately equal to the inverse of the FI
\begin{equation}
	\operatorname{Var}[\hat{\Theta}] \approx F(\Theta)^{-1}.
\end{equation}
In matrix form, $\Theta$ is a vector with N parameters. The FI is a matrix  $F_{i j}$
\begin{equation}\label{eq_fmdef1}
	F_{i j} =E\left[\frac{\partial l(\Theta ; x)}{\partial \Theta_i} \frac{\partial l(\Theta ; x)}{\partial \Theta_j}\right],
\end{equation}
and the inverse of FM is the covariance matrix $C_{i j}$,
\begin{equation}\label{ta}
	C_{i j}=F_{i j}^{-1}.
\end{equation}
\begin{table}[t]
	\caption{Notation of signal analysis}	\label{nsa}
	\begin{tabular}{|c|c|l|}
		\hline \text { variable } & \text { definition }&Dimension[E] \\
		\hline $h_{a b}(t, \vec{x})$ & metric perturbation&0 \\
		\hline $h_A(f, \hat{k})$ & Fourier amplitudes of metric perturbation&-1 \\
		\hline $S_h(f)$ &  PSD of GW background &-1 \\
		\hline $\Omega_{\mathrm{gw}}(f)$ & fractional energy density spectrum &0\\
		\hline $h_c(f)$ & characteristic strain &0\\
		\hline $h_{gw}(t)$ & detector response to GWs&0 \\
		\hline $n_{gw}(t)$ & detector response to noise&0 \\
		\hline $s_{gw}(t)$ & detector response &0 \\
		\hline$\tilde{h}_{gw}\left(f_n\right)$& discrete Fourier transform of $h_{gw}(t)$&0\\
		\hline $\tilde{h}_{gw}(f)$ & Fourier transform of $h_{gw}(t)$ &-1\\
		\hline $\tilde{n}_{gw}(f)$ & Fourier transform of $n_{gw}(t)$ &-1\\
		\hline $S_n(f)$ &  noise PSD for a single detector&-1 \\
		\hline $h_n(f)$ & characteristic strain noise amplitude for a single detector&0 \\
		\hline
		
	\end{tabular}
\end{table}

\section{Notation and corresponding dimension of signal analysis}\label{sec_nof}

TABLE~\ref{nsa} shows the notation of signal analysis and their dimension in natural units.

\bibliographystyle{JHEP}
\bibliography{axionref}

\end{document}